\documentclass[12pt,preprint]{aastex}

\newcommand{\Teff}{T_{\rm eff}}
\newcommand{\logg}{\log g}
\newcommand{\logeps}{\log \epsilon}
\newcommand{\vsini}{v \sin i}
\newcommand{\vrot}{v_{\rm rot}}
\renewcommand{\vr}{v_{\rm r}}
\newcommand{\kms}{\, {\rm km} \, {\rm s}^{-1}}
\newcommand{\K}{\, {\rm K}}

\newcommand{\I}{~{\scshape i}}
\newcommand{\II}{~{\scshape ii}}
\newcommand{\loggf}{\log g\!f}
\newcommand{\n}{\phantom{0}}

\newcommand{\etal}{{\it et al.}}

\shortauthors{Behr}
\shorttitle{Rotation Velocities of FHB stars}

\begin{document}

\title{Rotation Velocities of \\ Red and Blue Field Horizontal-Branch Stars}
\author{Bradford B. Behr}
\affil{McDonald Observatory \\ 1 University Station, C1400 \\ Austin TX 78712--0259 \email{bbb@astro.as.utexas.edu}}


\begin{abstract}

We present measurements of the projected stellar rotation velocities ($\vsini$) of a sample of 45 candidate field horizontal-branch (HB) stars spanning a wide range of effective temperature, from red HB stars with $\Teff \simeq 5000 \K$ to blue HB stars with $\Teff$ of $17000 \K$. Among the cooler blue HB stars ($\Teff = 7500$--$11500 \K$), we confirm prior studies showing that although a majority of stars rotate at $\vsini < 15 \kms$, there exists a subset of  ``fast rotators'' with $\vsini$ as high as 30--$35 \kms$. All but one of the red HB stars in our sample have $\vsini < 10 \kms$, and no analogous rotation bimodality is evident. We also identify a narrow-lined hot star ($\Teff \simeq 16000 \K$) with enhanced photospheric metal abundances and helium depletion, similar to the abundance patterns found among hot BHB stars in globular clusters, and four other stars that may also belong in this category. We discuss details of the spectral line fitting procedure that we use to deduce $\vsini$, and explore how measurements of field HB star rotation may shed light on the issue of HB star rotation in globular clusters.

\end{abstract}

\keywords{stars: horizontal-branch, stars: rotation, stars: abundances}


\section{Introduction and motivation}

The horizontal branch (HB) stars found in globular clusters exhibit a variety of photometric and spectroscopic characteristics which are not well explained by canonical models of stellar evolution. The relative number of red horizontal-branch (RHB) versus blue horizontal-branch (BHB) stars in a particular cluster is primarily a function of the cluster's metallicity, but sometimes two clusters with identical metallicities have very different HB color distributions, and it has proven difficult to conclusively identify the ``second parameters'' responsible for such differences \citep{fusipecci98}. Furthermore, many clusters show one or more ``gaps'' in the distribution of stars along their HB loci \citep{sosin97, ferraro98}, and regions where the stars are ``overluminous'' compared to models \citep{grundahl99}. Detailed spectroscopic observations of individual HB stars reveal additional peculiarities. The ``overluminous'' BHB stars are also found to have anomalously low surface gravities \citep{moehler95}, and photospheric abundance analyses of many of these same stars indicate enormous metal enhancements and helium depletion \citep{glaspey89, moehler99, behr99}. But perhaps the most perplexing characteristic of HB stars is the wide range of stellar rotation velocities that they exhibit.

The first comprehensive survey of globular cluster HB star rotation velocities was undertaken by Peterson and collaborators, and described in \cite{peterson83b, peterson85a, peterson85b} and \cite{peterson95}. In the five clusters studied, most of the blue HB stars had rotation velocities of 15 to $20 \kms$, but they found that about a third of the stars in cluster M13 were rotating twice as fast, with $\vsini$ as high as $40 \kms$. Such fast rotation is difficult to explain, given that these HB stars evolved from G-type main sequence stars, which are expected to have lost most of their primordial angular momentum to stellar winds during their main sequence lifetime. According to estimates by \cite{sills00}, the fastest observed rotation rates require complicated redistribution of angular momentum within the star as it evolves up the red giant branch (RGB), or addition of angular momentum from some external source, assuming that the main-sequence progenitors have $\vrot < 4 \kms$. Following the observational work of Peterson and collaborators, more fast-rotating BHB stars were subsequently found in several other globular clusters by \cite{cohen97}, \cite{behr00b}, and \cite{recioblanco02}. Of the 11 clusters measured to date, 5 have significant subpopulations of fast-rotating BHB stars, while the other have only the ``normal'' slow rotators.

Several different possible explanations for this anomalously fast rotation have been suggested. Angular momentum from a star's formation might be stored in a rapidly-rotating core \citep{peterson83a, peterson83b, pinsonneault91}, which gradually couples to the envelope only after the star reaches the HB. Tidal interactions with binary companions or absorption of Jovian planets in close orbits \citep{peterson83a, soker00, livio02, carney03} are other potential sources of ``excess'' angular momentum. An apparent correlation between high central cluster density and large populations of hotter BHB and EHB (extreme horizontal-branch) stars \citep{fusipecci93, buonanno97, testa01} suggests that cluster dynamics can directly influence the evolution of individual cluster stars, perhaps by imparting more internal angular momentum to stars that form in denser environments \citep{buonanno85}, so that the helium flash is delayed, and they lose more mass at the tip of the RGB and end up further to the blue end of the HB \citep{mengel76}. Under this latter scenario, we would expect higher-density clusters with blue tails to have more fast-rotating stars. The existing BHB rotation data does not support this connection --- fast HB rotation is seen in the low-density cluster M68, and no fast-rotating BHB stars are found in NGC~2808, a cluster with a long blue tail --- but some indirect influence is still a possibility.


One way to test these hypotheses regarding the role of the dynamical environment of clusters is to study the analogous field horizontal-branch (FHB) stars found in the metal-poor Galactic halo and thick disk, where the stellar density and the likelihood of close stellar encounters are considerably lower. If the rotation characteristics of the cluster BHB stars are somehow due to the dynamical environment of the cluster, then the field BHB stars should show a different distribution of rotation velocities. Candidate field BHB and RHB stars can be identified on the basis of their color, luminosity, and high proper motion, and those that are bright enough can then be observed at high spectral resolution in order to confirm that their gravities, metallicities, and rotation rates are those of old, evolved stars.

Measurements of stellar rotation among the field HB population have already been undertaken by several research groups. The first indication that some blue HB stars have unexpectedly high rotation rates came from \citet{peterson83a}, who presented $\vsini$ measurements of eight metal-poor field HB stars, as derived from synthesis fitting and Fourier-null analyses of the spectral profiles of metal absorption lines. Two of their eight stars had $\vsini$ values exceeding $30 \kms$, much like the fast rotators in globular clusters. \citet{kinman00} undertook a comprehensive study of known BHB stars, with a particular emphasis on chemical abundances, but also calculated approximate rotation velocities using Gaussian and synthesis fits to the strong Mg{\II} 4481 absorption line, finding a similar distribution of $\vsini$ values as previously reported. Only a few hot subdwarf B (sdB) stars, the field analogs of the hottest ``extreme HB'' stars in globular clusters, have been observed with sufficient detail to determine rotational velocities \citep{brassard01, rauch03}, and the intermediate-temperature BHB stars (10000--$15000 \K$) have been largely neglected because of the difficulties of distinguishing them from main-sequence A and B stars. At the cool end of the HB, \citet{carney03} present $\vsini$ values for field RHB stars, as determined from spectral cross-correlation with synthetic template spectra. They found that most RHB stars rotate more slowly than the BHB stars, with $\vsini < 10 \kms$, but also uncovered hints of a bimodal distribution of $\vsini$, as a few stars appeared to have somewhat higher $\vsini$.

This paper describes the results of an observational program intended to refine and extend the existing rotation measurements of field HB stars. We re-observed many of the stars analyzed in prior studies, so that a consistent measurement procedure could be applied to the whole sample. We also observed a number of other field stars which were considered HB candidates according to photometric criteria, in an attempt to expand the size of the field HB sample and its extent in parameter space. Section~2 details the selection of targets, the observations, and the spectral reduction pathway. Section~3 describes our analysis procedure, which uses photometric and spectroscopic data to determine stars' temperatures, gravities, microturbulence and rotation velocities, and chemical abundances. The results of these analyses are presented in Section~4, and in Section~5 we evaluate the accuracy of the rotational broadening measurements. In Section~6, we look for correlations between $\vsini$ and other stellar parameters. Section~7 briefly discusses the non-HB stars that turn up in our survey, and Section~8 summarizes the results of this study.


\section{Target selection, observations, and spectral reduction}

The BHB targets for this program were selected from a variety of sources. We included all of the ``confirmed'' BHB stars previously observed at high spectral resolution by \citet{peterson83a}, \citet{adelman87b}, and \citet{kinman00}, except for those that were too faint ($V > 11$) or located too far south. To expand the sample, we also selected candidate BHB targets from the high-latitude blue star lists of \citet{newell73} and \citet{newell76}, the FHB surveys of \cite{philip84}, \citet{kinman94}, \citet{gray96}, and \cite{wilhelm99}, and the kinematic surveys of \citet{beers95} and \citet{beers00}. We expected that the target lists from Newell, in particular, would yield many authentic BHB stars, in light of the photometric ``gaps'' that appear in their color-color diagrams, which seemed likely analogs to the gaps that appear in the HB color distributions of many globular clusters. We deliberately included several stars which had previously been ruled out as true HB stars, or were likely to be main-sequence A-star ``contaminants'', in order to test our discrimination protocols and confirm prior identifications.

Candidate RHB stars were drawn primarily from the extensive RHB abundance studies of \cite{taut97} and \cite{taut01}, with a few additional targets from \cite{strai81}. We specifically tried to avoid RR~Lyrae stars --- HB stars lying in the instability strip between 6000 and 7500 K --- as their variability would significantly complicate both the measurement and the interpretation of their rotation velocities. (A handful of the warmer RHB and cooler BHB candidates were subsequently found to have effective temperatures which put them within the instability strip, and were thus excluded from the rotation analysis, even if they were not specifically known to be variable.) We were also concerned about confusion among cool RHB stars, first-ascent RGB stars, and pre-AGB stars evolving redwards towards their second giant-branch ascent; identification of such potential contaminants is addressed later.

We must emphasize that because the target list for this study was drawn from so many different sources, it cannot be considered a complete (or even generally representative) sample. Each literature reference and catalog introduces its own influences, including biases towards higher proper motion, brighter absolute magnitudes, redder or bluer photometric colors, and such. Our final target list contains an indeterminate mixture of thin disk, thick disk, and halo populations, with different metallicity ranges, kinematic properties, and evolutionary types chosen according to different criteria, so no conclusions should be drawn regarding {\it e.g.} the relative numbers of stars in different evolutionary stages, or the field HB color morphology.

Most of the observations for this project were made using the Cassegrain Echelle Spectrograph \citep{mccarthy93} on the McDonald Observatory 2.1-meter Otto Struve Telescope. This instrument provides a nominal resolving power of $R \simeq 60000$ over a wavelength range of 1000 to 1500~\AA, depending on the central wavelength. Observations took place between April 1997 and March 2002, with many stars observed on multiple occasions in order to evaluate the precision of the subsequent analysis. We should note that these multiple observations are not suitable for looking for radial velocity variations among these stars --- although stable over short timescales, the spectrograph can be subject to thermal variations over the course of each night, and from night to night, resulting in spectral shifts of a few $\kms$. Binary star studies using this instrument usually take wavelength calibration exposures before and after every stellar exposure to avoid this potential problem, but as detection of stellar binarity was not a goal of this project, we chose not to incur the additional overhead this protocol would have required.

A handful of fainter stars ($V > 10$) were observed with the HIRES spectrograph \citep{vogt94} on the Keck~I telescope atop Mauna Kea, taking advantage of ``spare'' time during deep twilight. With the C1 slit decker, these spectra had $R = 45000$, with a much wider spectral coverage of $\sim 2500$~\AA. 

\begin{figure}
\epsscale{0.60}
\plotone{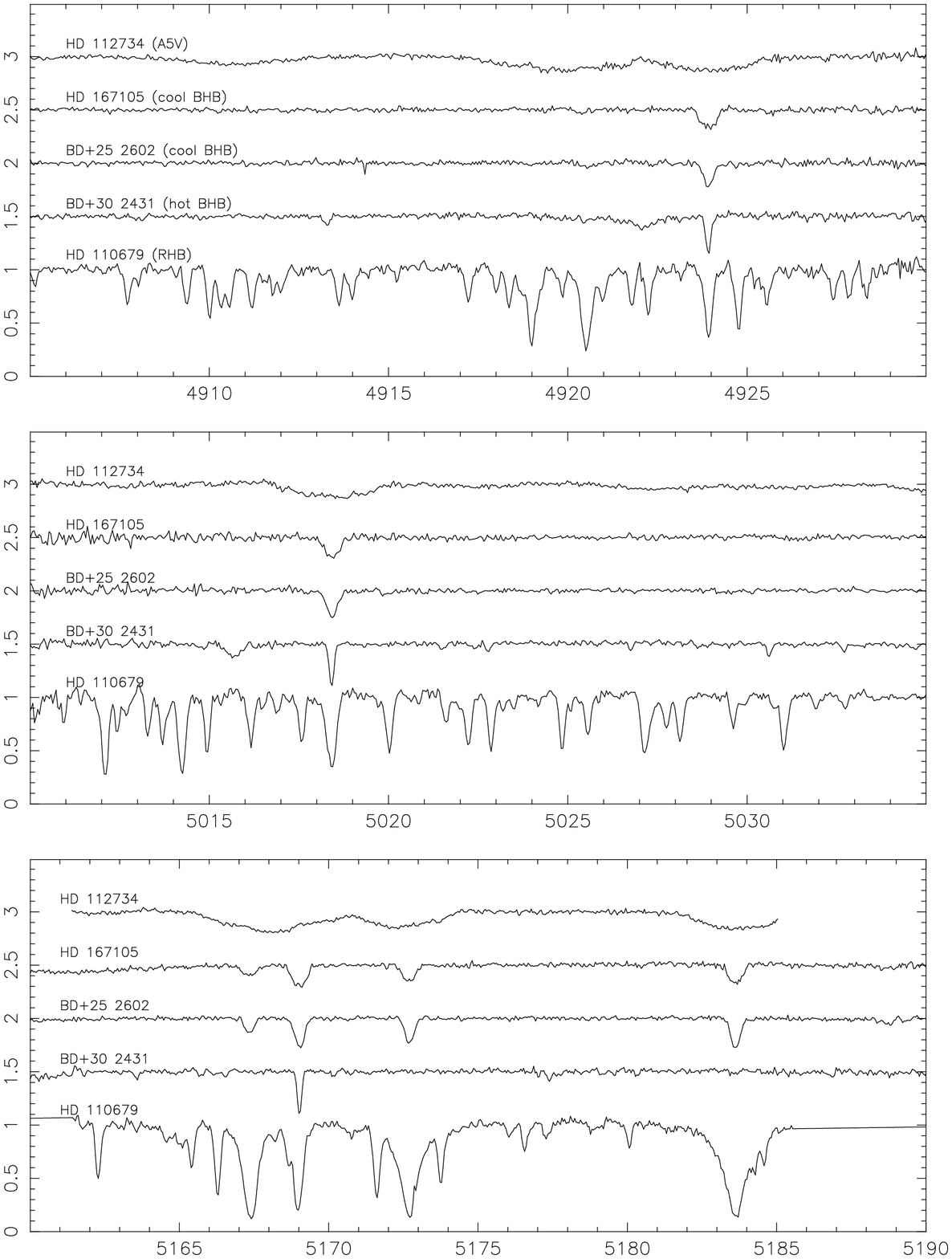}
\caption{Representative regions of the normalized spectra for five target stars, shifted to their respective rest frames, and offset vertically by multiples of 0.5 for clarity. \label{spectra}}
\end{figure}

Table~\ref{obs-table} summarizes the observations of target stars for this program, with dates and wavelength coverage of individual observations. For the McDonald Cassegrain Echelle (CE), three wavelength settings were used: ``blue'' covers 4180--4630~\AA, ``mid'' covers 4750--5490~\AA, and ``red'' covers 5460--6800~\AA. HIRES was usually set to span a ``blue'' range of 3890--5360~\AA, but a few observations were made with a ``red'' range of 5380--7830~\AA. The $S/N$ ratios (per pixel, not per resolution element) were computed from the reduced spectra by heavily smoothing the spectral regions identified as continuum, and then computing the rms deviation between the observed spectrum and the smoothed spectrum.
\clearpage
\begin{deluxetable}{lrllrr}
\tablecaption{Observations of FHB candidates. \label{obs-table}}
\tabletypesize{\scriptsize}
\tablewidth{0pt}
\tablehead{
			&		&			&		 	&$t_{\rm exp}\:$ 	&$S/N$ \\
star			&$V\n$ 	&civil date 	&instrument 	&(sec)			&per pixel\hspace{-6pt}}
\startdata
BD+00 0145	&10.6	&1998 Aug 26	&HIRES blue	&300	&130.8	\\
BD+01 0513	&10.7	&1998 Aug 26	&HIRES blue	&300	&82.7	\\*[-2pt]
			&		&2000 Dec 13	&CE red		&900	&33.5	\\*[-2pt]
			&		&2000 Dec 14	&CE red		&1200	&42.8	\\
BD+01 0514	&9.8		&2000 Dec 09	&CE blue		&648	&22.4	\\*[-2pt]
			&		&2000 Dec 09	&CE blue		&959	&20.7	\\*[-2pt]
			&		&2000 Dec 13	&CE red		&900	&49.3	\\
BD+01 0548	&10.7	&1998 Aug 26	&HIRES blue	&300	&137.6	\\*[-2pt]
			&		&1998 Aug 27	&HIRES blue	&300	&121.2	\\
BD+03 0740	&9.8		&2000 Oct 04	&CE blue		&1200	&26.6	\\*[-2pt]
			&		&2000 Dec 12	&CE blue		&867	&10.8	\\*[-2pt]
			&		&2000 Dec 14	&CE red		&1200	&66.7	\\*[-2pt]
			&		&2000 Dec 14	&CE red		&1200	&63.0	\\
BD+09 3223	&9.3		&2001 Mar 06	&CE mid		&900	&61.4	\\
BD+10 2495	&9.7		&2001 Jul 07	&CE mid		&900	&47.8	\\
BD+11 2998	&9.1		&2001 Mar 06	&CE mid		&900	&64.9	\\
BD+14 4757	&10.1	&1999 Aug 15	&HIRES blue	&120	&59.8	\\
BD+17 3248	&9.4		&1998 Apr 12	&CE blue		&2700	&21.7	\\*[-2pt]
			&		&1998 Apr 13	&CE blue		&1800	&46.3	\\*[-2pt]
			&		&1998 Apr 13	&CE blue		&1800	&46.7	\\
BD+17 4642	&9.1		&2000 Oct 10	&CE blue		&1200	&26.8	\\
BD+17 4708	&9.5		&1998 Oct 28	&CE blue		&1800	&44.9	\\*[-2pt]
			&		&1998 Oct 28	&CE blue		&1800	&44.7	\\*[-2pt]
			&		&1999 Aug 15	&HIRES blue	&60		&77.3	\\
BD+18 0153	&5.5		&2000 Dec 13	&CE red		&120	&161.8	\\
BD+18 2757	&9.8		&2001 Mar 11	&CE mid		&1500	&41.1	\\
BD+18 2890	&9.8		&2001 Jul 07	&CE mid		&900	&46.1	\\
BD+20 3004	&10.0	&2001 Mar 11	&CE mid		&1500	&40.5	\\*[-2pt]
			&		&2001 Mar 12	&CE red		&1200	&51.1	\\
BD+25 1981	&9.3		&2000 Dec 12	&CE blue		&1200	&18.8	\\*[-2pt]
			&		&2000 Dec 13	&CE red		&900	&76.4	\\*[-2pt]
			&		&2001 Mar 05	&CE mid		&900	&65.2	\\
BD+25 2436	&9.9		&2001 Mar 12	&CE red		&1200	&52.8	\\*[-2pt]
			&		&2001 Mar 13	&CE red		&1500	&67.7	\\
BD+25 2459	&9.6		&2001 Mar 12	&CE red		&1200	&69.4	\\
BD+25 2497	&10.3	&2001 Mar 06	&CE mid		&1200	&40.7	\\
BD+25 2602	&10.2	&2001 Jan 19	&CE blue		&1200	&44.6	\\*[-2pt]
			&		&2001 Jul 06	&CE mid		&1200	&66.0	\\
BD+27 2057	&9.5		&2001 Mar 12	&CE red		&900	&66.4	\\
BD+29 2231	&9.8		&2001 Mar 12	&CE red		&900	&62.5	\\
BD+29 2294	&9.5		&2001 Mar 10	&CE mid		&1200	&44.9	\\
BD+30 2338	&10.0	&2001 Mar 13	&CE red		&1200	&53.0	\\
BD+30 2355	&10.6	&2001 Mar 06	&CE mid		&1200	&39.7	\\*[-2pt]
			&		&2001 Jul 06	&CE mid		&900	&45.3	\\
BD+30 2431	&10.1	&2001 Mar 11	&CE mid		&1500	&38.3	\\*[-2pt]
			&		&2001 Mar 14	&CE red		&1200	&59.3	\\*[-2pt]
			&		&2001 Jul 06	&CE mid		&900	&54.6	\\
BD+32 2188	&10.7	&1998 Apr 20	&HIRES blue	&600	&230.2	\\*[-2pt]
			&		&1998 Apr 21	&HIRES blue	&300	&146.9	\\
BD+33 2171	&10.6	&1998 Apr 20	&HIRES blue	&300	&165.2	\\
BD+33 2642	&10.8	&2001 Mar 06	&CE mid		&1200	&35.7	\\*[-2pt]
			&		&2001 Mar 14	&CE red		&1500	&39.4	\\
BD+34 2371	&9.5		&2001 Mar 15	&CE red		&1200	&72.0	\\
BD+36 2242	&10.0	&2000 Jun 02	&HIRES red	&300	&232.5	\\*[-2pt]
			&		&2001 Jan 19	&CE blue		&1200	&52.7	\\
BD+36 2268	&10.3	&2001 Mar 06	&CE mid		&1200	&51.1	\\*[-2pt]
			&		&2001 Mar 12	&CE red		&1200	&48.4	\\
BD+36 2303	&9.6		&2001 Mar 10	&CE mid		&1200	&41.4	\\
BD+42 2309	&10.8	&1998 Apr 20	&HIRES blue	&600	&168.1	\\
BD+46 1998	&9.1		&2001 Mar 11	&CE mid		&450	&32.1	\\*[-2pt]
			&		&2001 Mar 14	&CE red		&450	&58.5	\\*[-2pt]
			&		&2001 Jul 07	&CE mid		&600	&64.0	\\
BD+49 2137	&10.7	&2001 Mar 15	&CE red		&1500	&42.8	\\
BD$-$02 0524	&10.3	&2000 Oct 04	&CE blue		&1200	&30.5	\\*[-2pt]
			&		&2000 Oct 10	&CE blue		&1200	&26.8	\\*[-2pt]
			&		&2000 Dec 14	&CE red		&1200	&59.4	\\
BD$-$07 0230	&11.1	&2000 Oct 04	&CE blue		&1200	&20.6	\\*[-2pt]
			&		&2000 Dec 14	&CE red		&1200	&36.8	\\
BD$-$12 2669	&10.3	&2000 Dec 14	&CE red		&1200	&51.0	\\*[-2pt]
			&		&2001 Mar 05	&CE mid		&900	&32.4	\\*[-2pt]
			&		&2001 Mar 06	&CE mid		&1200	&51.7	\\
BPS CS 22189--5	&14.2	&1997 Aug 03	&HIRES blue	&1500	&41.6	\\
BPS CS 22894--36	&14.8	&1997 Aug 01	&HIRES blue	&900	&16.5	\\
Feige 40		&11.1	&2001 Mar 06	&CE mid		&1200	&41.2	\\*[-2pt]
			&		&2001 Mar 09	&CE mid		&1200	&40.0	\\*[-2pt]
			&		&2001 Mar 13	&CE red		&1500	&34.5	\\
Feige 41		&11.0	&2001 Jan 19	&CE blue		&1200	&30.1	\\
Feige 84		&11.8	&2001 Mar 10	&CE mid		&1200	&15.2	\\*[-2pt]
			&		&2001 Mar 12	&CE red		&900	&21.9	\\
GCRV 63536	&11.1	&1998 Apr 20	&HIRES blue	&600	&129.1	\\
HD 97		&9.7		&1998 Oct 28	&CE blue		&1800	&30.5 	\\*[-2pt]
			&		&1998 Oct 28	&CE blue		&1800	&34.2 	\\*[-2pt]
			&		&2000 Dec 13	&CE red		&600	&54.7 	\\
HD 1112		&9.1		&2000 Oct 04	&CE blue		&1200	&54.3	\\*[-2pt]
			&		&2000 Oct 10	&CE blue		&600	&37.0	\\
HD 2857		&10.0	&2000 Oct 04	&CE blue		&1200	&28.9	\\*[-2pt]
			&		&2000 Oct 05	&CE blue		&1200	&24.7	\\
HD 2880		&8.6		&2000 Dec 10	&CE blue		&900	&24.3	\\
HD 6229		&8.6		&2000 Dec 10	&CE blue		&1200	&21.1	\\*[-2pt]
			&		&2000 Dec 13	&CE red		&180	&48.5	\\
HD 6461		&7.6		&2000 Dec 10	&CE blue		&600	&23.7	\\*[-2pt]
			&		&2000 Dec 13	&CE red		&120	&56.6	\\
HD 7374		&5.9		&2000 Dec 09	&CE blue		&180	&129.5	\\
HD 8376		&9.6		&1998 Oct 27	&CE blue		&950	&31.1	\\*[-2pt]
			&		&1998 Oct 27	&CE blue		&1200	&51.2	\\*[-2pt]
			& 		&2000 Dec 09	&CE blue		&1200	&61.1	\\
HD 13978		&9.6		&2000 Dec 09	&CE blue		&1200	&33.2	\\
HD 14829		&10.3	&2000 Dec 09	&CE blue		&1200	&30.3	\\
HD 24000		&8.8		&2000 Oct 10	&CE blue		&309	&20.8	\\
HD 24341		&7.8		&1998 Oct 28	&CE blue		&750	&58.7	\\*[-2pt]
			&		&1998 Oct 28	&CE blue		&750	&59.3	\\
HD 25532		&8.2		&2000 Oct 04	&CE blue		&600	&39.4	\\*[-2pt]
			&		&2000 Dec 13	&CE red		&240	&53.1	\\
HD 27295		&5.3		&2000 Dec 09	&CE blue		&120	&94.0	\\
HD 60778		&9.1		&2000 Dec 09	&CE blue		&1200	&38.6	\\*[-2pt]
			&		&2000 Dec 09	&CE blue		&1200	&48.0	\\*[-2pt]
			&		&2000 Dec 12	&CE blue		&1200	&26.8	\\*[-2pt]
			&		&2000 Dec 13	&CE red		&900	&68.5	\\*[-2pt]
			&		&2000 Dec 14	&CE red		&900	&70.1	\\*[-2pt]
			&		&2001 Mar 05	&CE mid		&900	&77.5	\\*[-2pt]
			&		&2001 Mar 06	&CE mid		&900	&83.1	\\
HD 63791		&7.9		&1998 Oct 28	&CE blue		&1200	&88.7	\\*[-2pt]
			&		&1998 Oct 28	&CE blue		&1200	&93.7	\\*[-2pt]
			&		&1998 Oct 31	&CE blue		&900	&54.3	\\*[-2pt]
			&		&1998 Oct 31	&CE blue		&900	&54.3	\\*[-2pt]
			&		&2000 Dec 13	&CE red		&450	&101.7	\\
HD 64488		&7.3		&2000 Dec 09	&CE blue		&120	&60.3	\\
HD 74721		&8.7		&1998 Apr 13	&CE blue		&1800	&81.2	\\*[-2pt]
			&		&1998 Apr 13	&CE blue		&1800	&89.7	\\
HD 76431		&9.2		&2001 Jan 19	&CE blue		&1200	&77.3	\\*[-2pt]
			&		&2001 Jan 19	&CE blue		&1200	&78.1	\\
HD 79452		&6.0		&2000 Dec 13	&CE red		&120	&124.0	\\*[-2pt]
			&		&2001 Jan 21	&CE red		&90		&130.8	\\
HD 79530		&9.6		&2000 Dec 12	&CE blue		&1200	&33.2	\\*[-2pt]
			&		&2000 Dec 13	&CE red		&900	&77.9	\\*[-2pt]
			&		&2001 Jan 21	&CE red		&900	&101.4	\\
HD 82590		&9.4		&2000 Dec 13	&CE red		&900	&67.3	\\
HD 83751		&8.7		&2001 Jan 19	&CE blue		&1200	&88.4	\\*[-2pt]
			&		&2001 Jan 19	&CE blue		&1200	&95.2	\\
HD 86986		&8.0		&2000 Dec 09	&CE blue		&300	&54.9	\\*[-2pt]
			&		&2000 Dec 09	&CE blue		&300	&55.7	\\
HD 87047		&9.7		&2000 Dec 09	&CE blue		&1200	&55.1	\\
HD 87112		&9.7		&1998 Apr 13	&CE blue		&1800	&70.2	\\*[-2pt]
			&		&1998 Apr 13	&CE blue		&1800	&68.2	\\
HD 93329		&8.8		&2000 Dec 09	&CE blue		&180	&22.4	\\*[-2pt]
			&		&2000 Dec 09	&CE blue		&450	&46.4	\\
HD 97560		&7.9		&2001 Jan 20	&CE red		&900	&101.7	\\*[-2pt]
			&		&2001 Jan 21	&CE red		&300	&93.3	\\
HD 100340	&10.1	&2001 Jan 19	&CE blue		&1200	&49.7	\\*[-2pt]
			&		&2001 Jan 21	&CE red		&600	&49.5	\\
HD 103376	&10.2	&2001 Jan 19	&CE blue		&1200	&49.4	\\
HD 105183	&11.0	&2001 Jan 19	&CE blue		&1200	&25.4	\\*[-2pt]
			&		&2001 Mar 06	&CE mid		&1200	&41.2	\\*[-2pt]
			&		&2001 Mar 12	&CE red		&1200	&33.9	\\
HD 105262	&7.1		&1998 Apr 13	&CE blue		&600	&115.4	\\*[-2pt]
			&		&1998 Apr 13	&CE blue		&600	&125.7	\\
HD 105546	&8.6		&1998 Apr 14	&CE blue		&1800	&39.7	\\*[-2pt]
			&		&1998 Apr 14	&CE blue		&1800	&45.7	\\*[-2pt]
			&		&2001 Mar 09	&CE mid		&600	&70.9	\\
HD 105944	&9.9		&2001 Mar 09	&CE mid		&900	&47.5	\\*[-2pt]
			&		&2001 Mar 15	&CE red		&900	&49.2	\\
HD 108577	&9.6		&1998 Apr 13	&CE blue		&1800	&46.2	\\*[-2pt]
			&		&1998 Apr 13	&CE blue		&1800	&40.8	\\*[-2pt]
			&		&2001 Mar 10	&CE mid		&1200	&44.6	\\
HD 109995	&7.6		&2001 Mar 07	&CE mid		&300	&41.8	\\*[-2pt]
			&		&2001 Mar 09	&CE mid		&300	&75.3	\\
HD 110679	&9.2		&2001 Mar 07	&CE mid		&1200	&31.7	\\*[-2pt]
			&		&2001 Mar 12	&CE red		&900	&86.7	\\
HD 110930	&9.8		&2001 Mar 10	&CE mid		&1200	&34.6	\\*[-2pt]
			&		&2001 Mar 12	&CE red		&600	&54.9	\\
HD 112030	&8.7		&2001 Mar 13	&CE red		&260	&30.1	\\*[-2pt]
			&		&2001 Mar 14	&CE red		&600	&59.6	\\
HD 112414	&9.4		&2001 Mar 06	&CE mid		&900	&65.1	\\
HD 112693	&9.4		&2001 Mar 11	&CE mid		&1500	&59.1	\\*[-2pt]
			&		&2001 Jul 07	&CE mid		&600	&57.1	\\
HD 112734	&7.0		&2001 Mar 07	&CE mid		&300	&47.8	\\*[-2pt]
			&		&2001 Mar 07	&CE mid		&600	&66.1	\\
HD 114839	&8.5		&2001 Mar 06	&CE mid		&600	&83.6	\\
HD 114930	&9.0		&2001 Mar 06	&CE mid		&900	&70.9	\\
HD 115520	&8.4		&1998 Apr 13	&CE blue		&1800	&86.3	\\
HD 115444	&9.0		&2001 Jul 08	&CE mid		&600	&71.3	\\
HD 117880	&9.1		&2001 Mar 07	&CE mid		&860	&23.8	\\*[-2pt]
			&		&2001 Mar 09	&CE mid		&900	&37.8	\\*[-2pt]
			&		&2001 Mar 12	&CE red		&600	&66.0	\\
HD 119516	&9.1		&2001 Mar 09	&CE mid		&900	&56.9	\\
HD 125924	&9.7		&2001 Mar 11	&CE mid		&1500	&47.6	\\*[-2pt]
			&		&2001 Mar 14	&CE red		&900	&52.3	\\
HD 128801	&8.8		&2001 Mar 09	&CE mid		&900	&49.3	\\*[-2pt]
			&		&2001 Mar 12	&CE red		&450	&77.7	\\*[-2pt]
			&		&2001 Jul 05	&CE blue		&900	&59.6	\\*[-2pt]
			&		&2001 Jul 06	&CE mid		&1200	&90.9	\\
HD 130156	&9.3		&2001 Jul 06	&CE mid		&600	&54.8	\\
HD 135485	&8.1		&2001 Mar 09	&CE mid		&600	&67.4	\\
HD 137569	&7.9		&2001 Mar 09	&CE mid		&600	&66.1	\\*[-2pt]
			&		&2001 Mar 14	&CE red		&300	&74.4	\\
HD 143459	&5.5		&2001 Mar 09	&CE mid		&60		&84.4	\\
HD 145293	&10.0	&2001 Mar 09	&CE mid		&1200	&30.0	\\*[-2pt]
			&		&2001 Mar 09	&CE mid		&1200	&25.7	\\*[-2pt]
			&		&2001 Mar 14	&CE red		&1500	&50.3	\\
HD 159061	&8.4		&2001 Jul 05	&CE blue		&1200	&39.8	\\
HD 161770	&9.7		&2001 Jul 07	&CE mid		&900	&39.7	\\*[-2pt]
			&		&2001 Jul 08	&CE mid		&1200	&58.0	\\
HD 161817	&7.0		&2001 Jul 05	&CE blue		&300	&44.7	\\*[-2pt]
			&		&2001 Jul 05	&CE blue		&600	&51.3	\\*[-2pt]
			&		&2001 Jul 06	&CE mid		&300	&132.5	\\
HD 166161	&8.1		&2001 Jul 07	&CE mid		&600	&83.8	\\
HD 167105	&8.9		&2001 Jul 05	&CE blue		&1800	&63.4	\\*[-2pt]
			&		&2001 Jul 06	&CE mid		&300	&59.1	\\
HD 167768	&6.0		&2001 Jul 07	&CE mid		&60		&56.8	\\
HD 170357	&8.3		&2001 Jul 08	&CE mid		&300	&63.2	\\
HD 175305	&7.2		&2001 Jul 06	&CE mid		&60		&39.7	\\
HD 180903	&9.6		&2001 Jul 06	&CE mid		&900	&52.2	\\
HD 184266	&7.6		&2001 Jul 06	&CE mid		&120	&39.2	\\*[-2pt]
			&		&2001 Jul 06	&CE mid		&240	&63.2	\\
HD 195636	&9.5		&2001 Jul 06	&CE mid		&900	&55.8	\\
HD 199854	&8.9		&2000 Oct 05	&CE blue		&900	&21.9	\\*[-2pt]
			&		&2000 Oct 11	&CE blue		&1200	&33.2	\\
HD 202573	&7.0		&2001 Jul 06	&CE mid		&180	&54.1	\\
HD 203563	&8.2		&2000 Jun 02	&HIRES red	&280	&224.1	\\*[-2pt]
			&		&2000 Oct 05	&CE blue		&1200	&76.4	\\*[-2pt]
			&		&2000 Dec 13	&CE red		&300	&74.5	\\*[-2pt]
			&		&2000 Dec 13	&CE red		&600	&118.3	\\*[-2pt]
			&		&2001 Jul 05	&CE blue		&900	&70.8	\\*[-2pt]
			&		&2001 Jul 06	&CE mid		&450	&72.5	\\*[-2pt]
			&		&2001 Jul 07	&CE mid		&600	&95.8	\\*[-2pt]
			&		&2001 Jul 08	&CE mid		&900	&112.0	\\
HD 203854	&9.2		&1998 Oct 28	&CE blue		&1800	&94.4	\\*[-2pt]
			&		&1998 Oct 28	&CE blue		&1800	&88.7	\\*[-2pt]
			&		&2000 Oct 04	&CE blue		&1200	&57.1	\\
HD 208110	&6.2		&2001 Jul 06	&CE mid		&30		&45.0	\\
HD 210822	&8.2		&2000 Oct 11	&CE blue		&900	&68.2	\\
HD 213781	&9.0		&2000 Oct 04	&CE blue		&1200	&49.3	\\*[-2pt]
			&		&2000 Oct 11	&CE blue		&1200	&51.0	\\
HD 214994	&4.8		&1998 Oct 28	&CE blue		&180	&137.5	\\*[-2pt]
			&		&1998 Oct 28	&CE blue		&180	&137.6	\\
HD 217515	&9.4		&2000 Oct 04	&CE blue		&1200	&32.8	\\
HD 218790	&7.2		&2000 Dec 09	&CE blue		&600	&37.1	\\*[-2pt]
			&		&2000 Dec 13	&CE red		&120	&71.5	\\
HD 220787	&8.3		&2000 Oct 04	&CE blue		&600	&55.7	\\*[-2pt]
			&		&2000 Oct 10	&CE blue		&600	&56.1	\\
HD 229274	&9.1		&2001 Jul 06	&CE mid		&700	&61.9	\\
HD 233532	&10.1	&2000 Dec 14	&CE red		&1200	&61.1	\\
HD 233622	&10.0	&2000 Dec 09	&CE blue		&1200	&53.8	\\
HD 233666	&9.3		&1998 Apr 14	&CE blue		&1800	&41.9	\\*[-2pt]
			&		&1998 Apr 14	&CE blue		&1800	&42.1	\\*[-2pt]
			&		&2000 Dec 13	&CE red		&900	&79.4	\\
HD 252940	&9.1		&2000 Dec 09	&CE blue		&1200	&31.7	\\*[-2pt]
			&		&2000 Dec 09	&CE blue		&1200	&46.2	\\*[-2pt]
			&		&2000 Dec 13	&CE red		&900	&85.9	\\
HZ 27		&10.4	&2001 Mar 06	&CE mid		&1200	&14.6	\\
PG 0122+214	&* 10.8	&1998 Aug 26	&HIRES blue	&300	&55.3	\\
PG 0314+146	&* 10.6	&2000 Dec 14	&CE red		&1200	&23.8	\\
PG 0823+499	&* 12.4	&2000 Dec 09	&CE blue		&1800	&20.6	\\*[-2pt]
			&		&2000 Dec 09	&CE blue		&1800	&22.3	\\
PG 0855+294	&* 11.0	&2000 Dec 14	&CE red		&1200	&15.7	\\*[-2pt]
			&		&2001 Mar 06	&CE mid		&1200	&22.5	\\*[-2pt]
			&		&2001 Mar 09	&CE mid		&1200	&21.9	\\
PG 1205+228	&* 9.8	&2001 Mar 09	&CE mid		&900	&35.1	\\*[-2pt]
			&		&2001 Mar 11	&CE mid		&1500	&29.4	\\*[-2pt]
			&		&2001 Mar 12	&CE red		&900	&36.7	\\
PG 1530+212	&* 9.9	&2001 Mar 10	&CE mid		&1200	&27.4	\\*[-2pt]
			&		&2001 Mar 12	&CE red		&1200	&33.8	\\*[-2pt]
			&		&2001 Mar 14	&CE red		&1200	&33.2	\\*[-2pt]
			&		&2001 Jul 05	&CE blue		&1800	&20.4	\\*[-2pt]
			&		&2001 Jul 06	&CE mid		&900	&41.7	\\
PG 2219+094	&* 9.4	&1999 Aug 15	&HIRES blue	&120	&33.0	\\*[-2pt]
			&		&2000 Jun 02	&HIRES red	&300	&97.7	\\
PG 2345+241	&* 11.6	&2000 Oct 10	&CE blue		&1200	&10.2	\\
PHL 25		&12.0	&2000 Oct 11	&CE blue		&1200	&9.2		\\
PHL 3275		&11.2	&2000 Dec 14	&CE red		&1200	&34.3	\\
\enddata
\tablecomments{* $V$ magnitude not available in literature, $B$ magnitude listed instead}
\end{deluxetable}
\clearpage
To reduce the spectral images to arrays of one-dimensional spectra, we employed a suite of routines developed by \cite{mccarthy90} for the FIGARO data analysis package \citep{shortridge93}. CCD frames were first bias-subtracted, using the overscan region from each frame. A master normalized flatfield frame was constructed from several different exposures of the spectrograph's internal incandescent lamps, median-filtered to remove cosmic ray hits, and appropriately weighted to provide uniform response from order to order. Cosmic ray hits on the data frames were identified and removed by hand. Each spectral order was traced by a 10th-order polynomial, and the spectrum extracted via direct pixel summation. Sky continuum and scattered light were extracted from windows immediately above and below the stellar trace. Both spectrographs use an internal thorium-argon arc lamp for wavelength calibration, and residuals on the polynomial fit for the wavelength solution for each order averaged 10~m\AA\ or less. The thorium-argon spectra were also used to construct an instrumental broadening profile for each night of observations. We normalized each order to unity by fitting a 6th- to 8th-order polynomial to likely continuum regions, as selected by eye. (Additional adjustments to the continuum are made later, during spectral synthesis comparisons, using spectral regions that are clearly line-free in the synthetic spectra.) We did {\it not} merge separate spectral orders or multiple observations of the same star, as our analysis codes are designed to handle multiple instances of the same spectral feature. Representative spectral regions from five different target stars are shown in Figure~\ref{spectra} to illustrate the appearance of the final spectra.

			
\section{Photometric and spectroscopic analysis}

For each of our target stars, we wish to determine several photospheric parameters. Effective temperature $\Teff$, surface gravity $\logg$, and metallicity [Fe/H] can be used to deduce the star's evolutionary status, and verify that it belongs to the HB, and not the RGB, the asymptotic giant branch (AGB), or the main sequence (MS). The magnitude of the microturbulence velocity $\xi$ must also be determined in order to calculate detailed chemical abundances $\logeps$ for each atomic species, both for the overall metallicity of the star, and to look for abundance peculiarities of interest. We employed a combination of photometric and spectroscopic techniques to constrain $\Teff$ and $\logg$, and then used spectral synthesis techniques to iteratively solve for $\xi$, $\vsini$, and $\logeps$.

Photometric colors were drawn from the literature, using the SIMBAD database and the General Catalogue of Photometric Data \citep{mermilliod97}. Nearly all of the target stars had $B-V$ and $U-B$ colors, from which we computed the reddening-free parameter $Q = (U-B) - 0.72 (B-V)$. A majority of the stars had also been observed in Str\"omgren $ubvy$ colors, which yielded the reddening-free Balmer jump index $[c_1] = c_1 - 0.20 (b-y)$ and reddening-free metallicity index $[m_1] = m_1 + 0.32 (b-y)$. The compilation of \cite{hauck98} provided most of these Str\"omgren data, with supplementary observations of hot subdwarfs from \cite{moehler90}, \cite{wesemael92}, and \cite{mooney00}. (We note that the Str\"omgren colors listed by Simbad for the star PHL~3275 appear to be erroneously duplicated from a cooler nearby star.) About 20\% of the stars also had Geneva photometry available, primarily from \cite{rufener76}, which we converted into reddening-free Geneva indices $X$ and $Y$ \citep{cramer84}. For a handful of stars, no photometric data could be found in the literature, so we relied exclusively on the spectroscopic techniques described below.

\begin{figure}
\epsscale{0.90}
\plotone{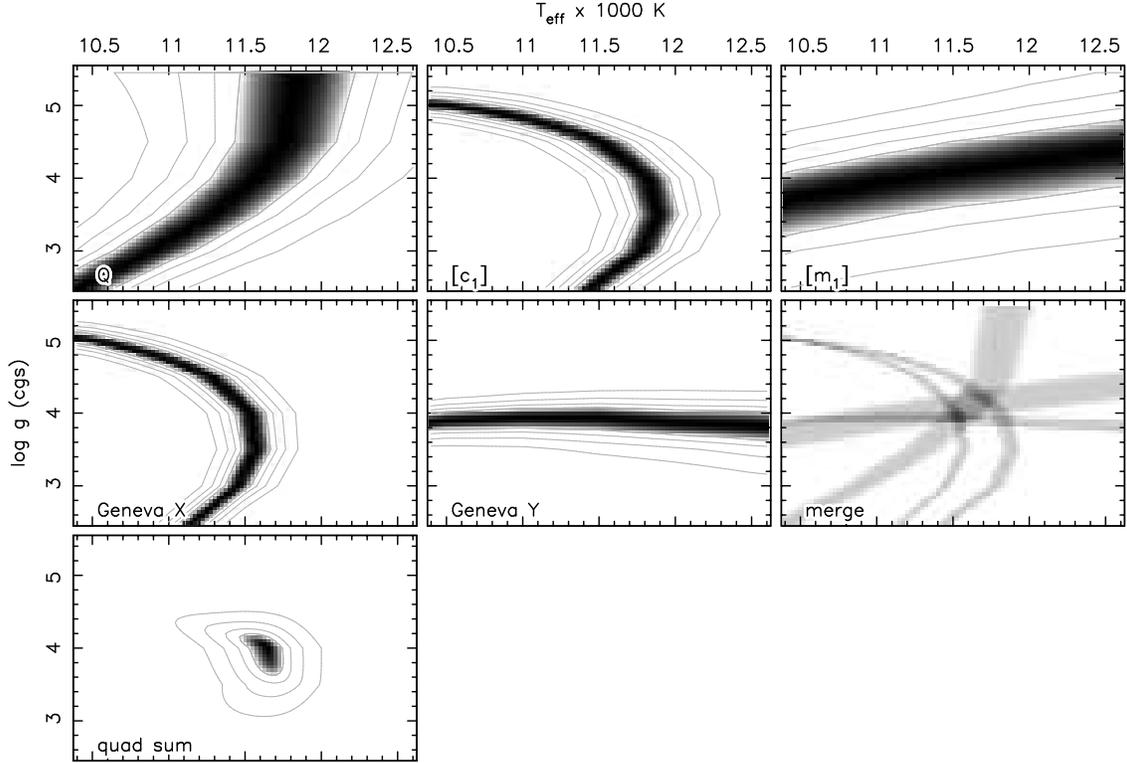}
\caption{Photometric quality-of-fit $z$ as a function of $\Teff$ and $\logg$. The greyscale runs from $z = 0$ (black) to $z = 1$, and contour lines are plotted for $z = 1, 2, 4, 8$. Each reddening-free color index (Johnson $Q$, Str\"omgren $[c_1]$ and $[m_1]$, and Geneva $X$ and $Y$) defines a different swath through the $(\Teff, \logg)$ plane. (The angular shape of some of the contour lines is due to the coarse resolution of the grid of synthetic colors.) In the ``merge'' panel, the swaths are overplotted to show that they converge near a mutually-agreeable solution. The quadrature sum of all the $z$ maps indicates a solution of $\Teff \simeq 11600 \K, \logg \simeq 4.0$ for this star. \label{phot-zmaps}}
\end{figure}

We compared these observed photometric colors to synthetic colors derived from grids of ATLAS9 model atmospheres \citep{kurucz93, kurucz03}. We initially assumed a metallicity of [Fe/H]$= -1.0$ for each star, and interpolated synthetic colors for each color index over a grid of $\Teff$ and $\logg$ values. At each point in the $(\Teff, \logg)$ plane, we calculated a quality-of-agreement parameter $z$ (in essence, a $\chi^2$ measure with one degree of freedom) by comparing the synthetic photometric color $C_{\rm syn}$ and the observed color $C_{\rm obs} \pm \sigma(C_{\rm obs})$~:
\begin{equation}
z = ((C_{\rm obs}-C_{\rm syn})/\sigma(C_{\rm obs}))^2
\end{equation}
such that $z=0$ where $C_{\rm syn}$ and $C_{\rm obs}$ are in perfect agreement, $z=1$ where $C_{\rm syn} = C_{\rm obs} \pm 1\sigma$, and $z \gg 1$ as $C_{\rm syn}$ and $C_{\rm obs}$ differ significantly. Thus, each observed photometric color defines a different zone or swath across the $(\Teff, \logg)$ plane within which the synthetic and observed colors agree. If all the different zones converge on a single small region, this indicates a likely photometric solution for the star's temperature and gravity. Figure~\ref{phot-zmaps} illustrates this approach, with the $z$-maps for six different color indices for one star plotted with greyscales and contours. The $z$ values at each map point are added quadratically and then normalized to yield a ``summation'' map, which shows the values and error intervals over which $\Teff$ and $\logg$ yield the best photometric agreement. 

This approach is potentially subject to several sources of systematic error. Most photometric color indices are sensitive to the photospheric metal abundance, so our initial assumption of [Fe/H]$=-1$ may result in errors of several hundred K in $\Teff$ and 0.5 dex or more in $\logg$ if the star has a significantly different metallicity. The crude preliminary $\Teff/\logg$ solution is still adequate, however, for an initial chemical abundance analysis, which we use to update the adopted metallicity for the subsequent iteration, so this error source is swiftly minimized. Interstellar reddening and extinction can also complicate the interpretation of photometric colors, but by using reddening-free color indices like $Q$, $[c_1]$, and $[m_1]$, we are able to avoid this issue. Probably the largest potential pitfall in this photometric $z$-map technique is the influence of systematic errors in the synthetic photometry. Even the most advanced model atmosphere models are not perfect, and resulting grids of synthetic photometry may require ``tweaking'' to bring them into agreement with standard stars of ``known'' $\Teff$ and $\logg$, as described by \cite{moon85}. We have applied no such adjustments or corrections to the Kurucz synthetic color data used in our analysis, so the photometrically-derived parameters may suffer from small systematic offsets. For most of our stars, however, we find that the $\Teff$ and $\logg$ derived from photometry agree well with the $\Teff$ and $\logg$ constraints placed by spectroscopic analysis, as described below. 

Absolute magnitudes for the target stars would have permitted us to further constrain $\logg$, as was done for globular cluster BHB stars by \cite{behr03}, but many of our field star targets are too far away to have accurate geometric parallaxes from Hipparcos or ground-based measurements, so this approach was not used. 

For analysis of the stellar spectra, we employed the LINFOR spectral synthesis code \citep{lemke97}, along with model atmospheres computed with ATLAS9, and atomic parameters drawn from the linelist compilation of \cite{hirata94}. (In a few cases, the $\loggf$ values from Hirata yielded wildly discordant results for lines of the same species, so we used $\loggf$ values from the VALD database \citep{piskunov95, ryabchikova99, kupka99} instead.) Instead of measuring the equivalent width of each metal absorption line in a spectrum, and deriving chemical abundances from those numbers, we directly compare an observed spectrum to a series of synthetic spectra, varying one or more synthesis parameters at a time in order to find an optimum fit between theory and observation. 

Solving for abundances ($\logeps$) for each relevant chemical species is relatively straightforward. For each species, we scan through a range of values of $\logeps$, calculating a synthetic spectrum for each of the lines of that species, convolving them by the empirically-determined instrumental profile, mapping to the wavelength bins of the observed spectrum, and calculating the rms deviation between the observed and synthetic line spectra. The minimum point in the rms curve indicates the best-fit value of $\logeps$, and a $1\sigma$ confidence interval for the species abundance is delineated by the values of $\logeps$ for which the rms increases by an amount corresponding to $\chi^2 = \chi_{\rm min}^2 + 1$. This quality-of-fit curve is most conveniently represented by a quantity
\begin{equation}
z = \sqrt{N_{\rm points}/2} \left( {{\rm rms}^2 \over {\rm rms}^2_{\rm min}} - 1 \right)
\end{equation}
which is analogous to the $z$ defined for the photometric analysis above, such that $z = 0$ at the best-fit value of $\logeps$, and the values where $z = 1$ represent the $1\sigma$ confidence interval. 

The radial velocity $\vr$ is determined in a similar fashion, by stepping through a sequence of values for $\vr$, and comparing the Doppler-shifted synthetic spectrum to the observed spectrum. The minimum in the $z$ curve marks the best-fit value of $\vr$, with error bars set by the points where $z = 1$. If a star is observed multiple times, we compute a separate $\vr$ for each observation, since the rotation and orbital motion of the Earth will induce a $\vr$ offset that varies from observation to observation. These values for observed telescope-centric $\vr$ are later shifted to the heliocentric reference frame.

To solve for $\xi$ or $\vsini$, we must take a two-dimensional slice through the multi-dimensional parameter space. The program adopts a value for $\xi$, and then adjusts $\logeps$ for each species to find the optimum fit for all the lines of that species. If the selected value of $\xi$ is close to the true value, then each of the spectral lines of a species will achieve a good fit at nearly the same $\logeps$, and the total rms deviation between observation and theory will be small. If, however, $\xi$ is too small or too large, then the weak lines and strong lines of a species will attain their best fits at different values of $\logeps$, and the total rms for that ensemble of lines will not reach as small a minimum value. Thus, as the program scans through a range of values for $\xi$, with $\logeps$ as free parameters, we can determine the value of $\xi$ that gives the best global rms, and define a confidence interval for $\xi$ over which $\chi^2 < \chi_{\rm min}^2 + 1$. This approach serves a similar purpose as the common technique of adjusting $\xi$ such that the $\logeps$ values derived from measured equivalent widths of multiple lines of the same species show no slope or trend when plotted as a function of equivalent width.

Once we have calculated a value for the microturbulent velocity, we adopt the same value for the macroturbulence parameter, since our spectra do not have sufficient resolution to disentangle the line broadening due to macroturbulence from that of the star's rotation. It is a gross oversimplification to assume that the velocity fields in a stellar atmosphere are the same at both small and large length scales, but this procedure does reflect the expectation that the stable radiative atmospheres of the hotter stars should have $\xi = v_{\rm macro} \simeq 0 \kms$, while $\xi > 0 \kms$ and $v_{\rm macro} > 0$ for the convective envelopes of the cooler stars. 

\begin{figure}
\epsscale{0.90}
\plottwo{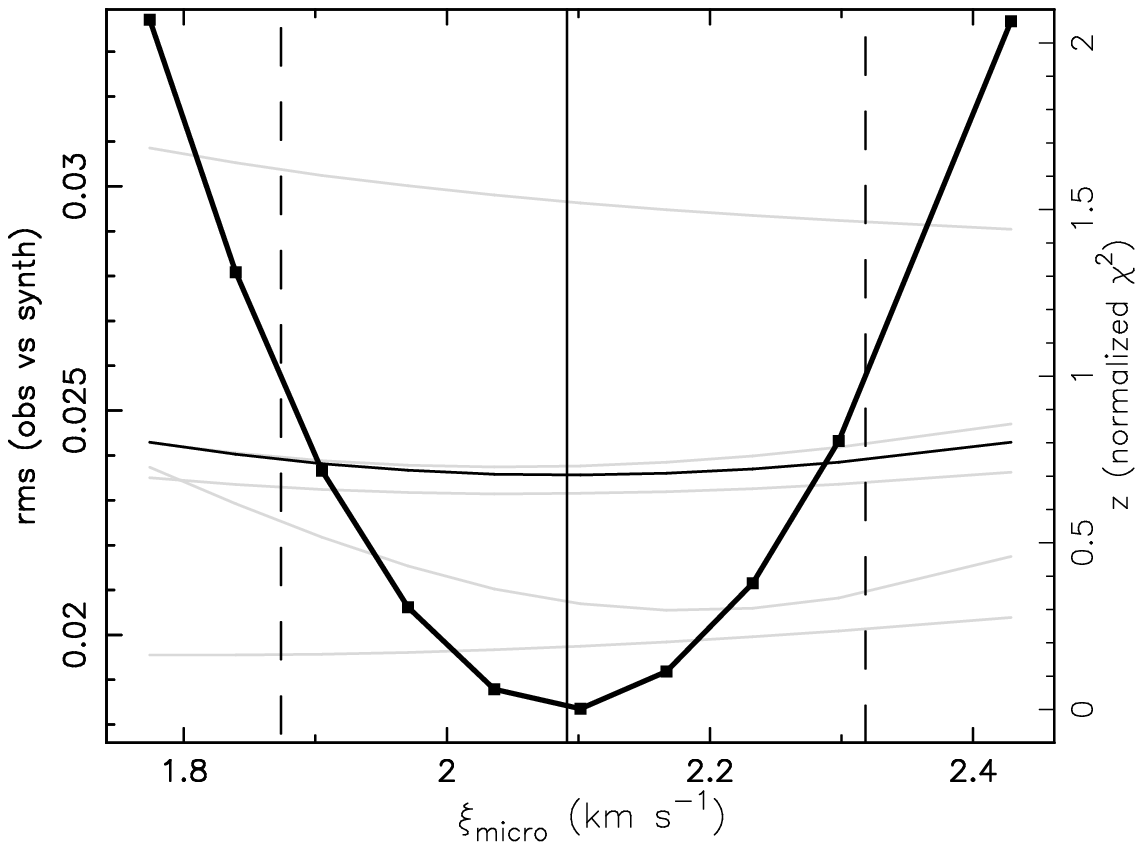}{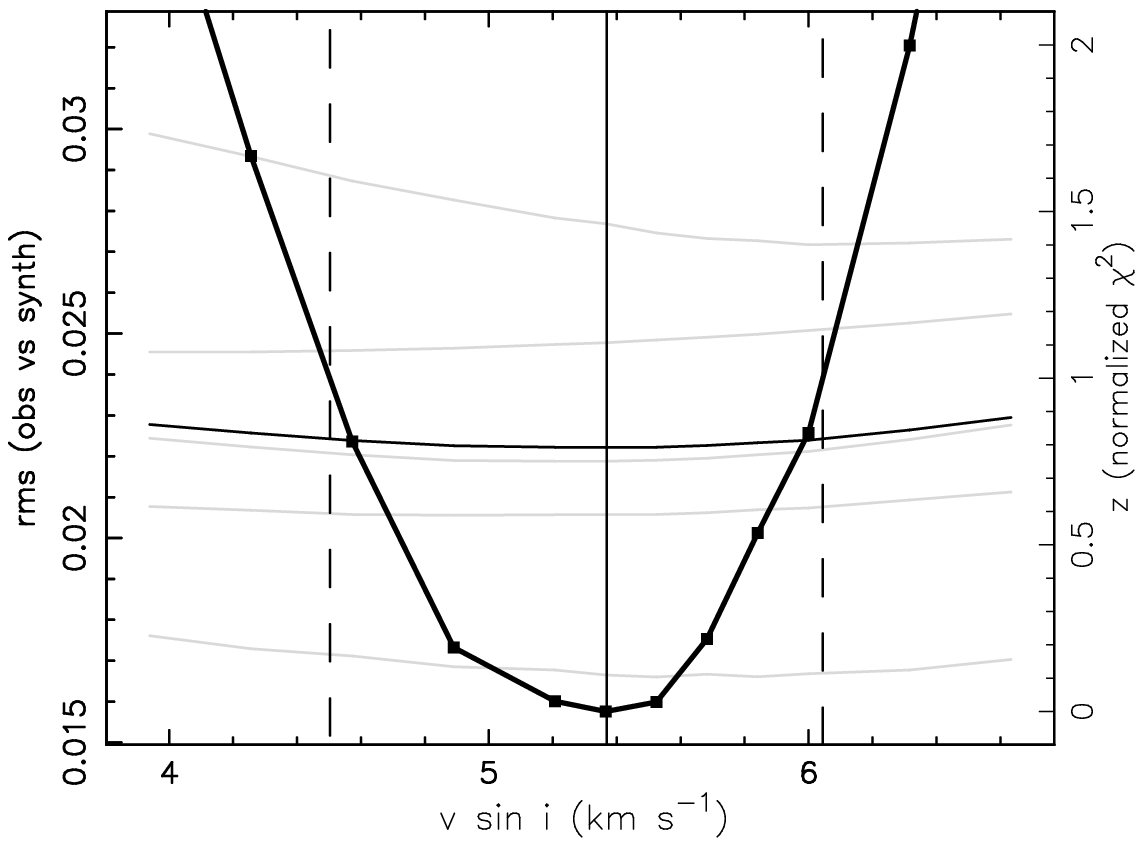}
\caption{Quality-of-fit curves to determine microturbulence $\xi$ (left panel) and $\vsini$ (right panel) from spectral line fitting. We step through a range of values for $\xi$ or $\vsini$, and at each step, we permit $\logeps$ for each species to vary freely, to find the best fit (smallest rms deviation) between synthetic and observed metal line profiles. The grey curves show the variation in rms for each of five different metal species, and the thin black curve shows the total rms deviation (the weighted sum across all five species) as a function of $\xi$ or $\vsini$. The heavy black curve shows the value of $z$, a normalized $\chi^2$ measure, which is computed from the total rms value as described in the text. The point where the total rms reaches a minimum indicates the value of $\xi$ or $\vsini$ that gives the best global solution (solid vertical line) , and the ordinate values where $z = 1$, corresponding to $\chi^2 = \chi_{\rm min}^2 + 1$, determine the $\pm 1\sigma$ error bars for $\xi$ or $\vsini$ (dashed vertical lines). \label{z-xi-vsini}}
\end{figure}

To solve for $\vsini$, we do a similar sort of two-dimensional scan as for $\xi$. Values of $\vsini$ that are too large or too small will result in mismatches between the synthetic and observed line profile shapes, and thus larger rms values, even when $\logeps$ is permitted to vary as a free parameter. Figure~\ref{z-xi-vsini} shows the quality-of-fit curves from scans of $\xi$ and $\vsini$ from a representative spectral analysis. Line profiles with $\vsini < 4 \kms$ are clearly excluded by the spectral fitting, showing how we can measure rotation velocities which are comparable, or even smaller, than the velocity resolution of the spectrograph. Potential systematic effects in determining $\vsini$ are discussed in Section~5 below.

\begin{figure}
\epsscale{0.80}
\plotone{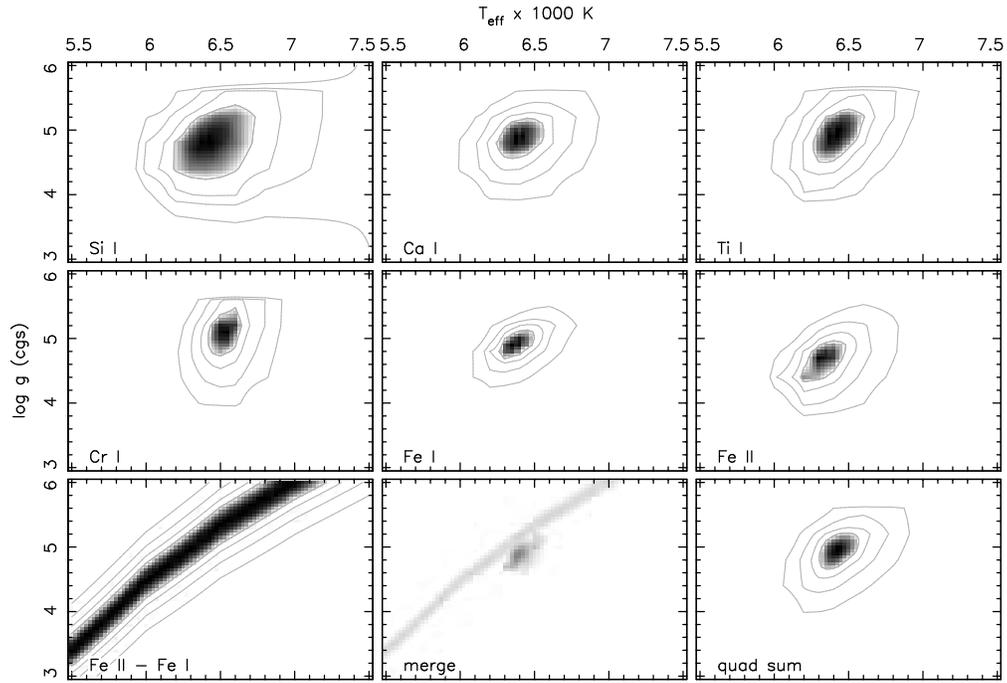}
\caption{Maps of the normalized $\chi^2$ parameter $z$ over the $(\Teff, \logg)$ plane, as derived from spectral synthesis fits for each of six chemical species --- Si{\I}, Ca{\I}, Ti{\I}, Cr{\I}, Fe{\I}, and Fe{\II} --- and the ionization balance between Fe{\I} and Fe{\II} (lower left panel). The greyscale runs from $z = 0$ (black) to $z = 1$, and contour lines are plotted for $z = 1, 2, 4, 8$. The quadrature sum of all the maps gives a global best-fit solution of $\Teff \simeq 6400$~K, $\logg \simeq 5.0$ for this star. \label{tg-ionz-zmaps}}
\end{figure}

This process can be extended to three dimensions in order to constrain $\Teff$ and $\logg$. If we adopt the correct temperature and gravity for our analysis, then all the lines of a given species will reach a good fit at (or near) the same value of $\logeps$, and we will compute a small value for the total rms. The different spectral lines of a species, with different excitation potentials $\chi$, will often respond differently to changes in $\Teff$ and $\logg$, however, so if we use incorrect values for $\Teff$ and $\logg$, we will not be able to get as good a fit, no matter how we vary $\logeps$. Thus we can search through the 2-dimensional temperature vs. gravity plane, and at each $(\Teff, \logg)$ point, let $\logeps$ vary to find the minimum local rms. Once the entire plane (within user-defined boundaries) has been well-sampled, the global minimum value of the rms can be used to calculate $z$ at each point. Plotting $z$ as a function of $\Teff$ and $\logg$, we can create a map of likely solution zones, similar to the maps derived from photometry, as shown in Figure~\ref{tg-ionz-zmaps}. The close agreement among the maps from several different species provides an indication that the solution is robust, and that none of the single-species synthesis maps have been unduly influenced by incorrect values for individual lines' $\loggf$ values or other atomic transition parameters.

Additional limits on temperature and gravity can be set by using iron ionization equilibrium, {\it i.e.} assuming that $\logeps($Fe\I) should equal $\logeps($Fe\II). In practice, we must modify this assumption slightly to account for discrepancies between our models and the actual stellar photospheres. Our version of LINFOR spectral synthesis code assumes full local thermodynamic equilibrium (LTE) for computing ionization populations, but the photospheres of metal-poor stars are subject to various non-LTE effects, such as ``overionization'' of neutral species by the UV radiation field. In order to account for this limitation in the models, we assume a non-LTE offset, $\Delta\logeps_{\rm nLTE} = \logeps_{\rm LTE}({\rm II}) - \logeps_{\rm LTE}({\rm I})$, which is primarily a function of metallicity. We use $\Delta\logeps_{\rm nLTE} = -0.1 \times {}$[Fe\II/H], as estimated from Figure~9 of \cite{thevenin99}. Such a simple relation is undoubtedly an oversimplification, as the atomic level populations also depend upon temperature, pressure, and many additional details of atmospheric structure and radiative transfer, but this crude non-LTE adjustment is sufficient to determine whether a particular choice of $(\Teff, \logg)$ is reasonable. (In contrast, \cite{kinman00} find a mean offset $\langle\,$[Fe\I/H]${} - {}$[Fe\II/H]$\,\rangle = 0.01 \pm 0.01$ for their BHB sample, so $\Delta\logeps_{\rm nLTE} = 0$ might be a better choice for these stars.) We scan through the $(\Teff, \logg)$ plane, computing $\logeps$ for both neutral and singly-ionized iron, and calculating yet another version of the quality-of-fit parameter
\begin{equation}
z = \left( {\logeps({\rm Fe~II}) - \logeps({\rm Fe~I}) - \Delta\logeps_{\rm nLTE} \over \sigma_{\rm II} + \sigma_{\rm I}} \right)^2
\end{equation}
which is defined so that $z=0$ when $\logeps($Fe\I) and $\logeps($Fe\II) differ by the predicted non-LTE offset. Figure~\ref{tg-ionz-zmaps} includes the $z$-map for iron ionization in the lower left panel, which illustrates reasonable (but not perfect) agreement with the single-species $\Teff$ vs $\logg$ maps.

All of the individual photometric $z$-maps for a particular star are merged into one, as illustrated in Figure~\ref{phot-zmaps}, and all individual single-species synthesis $z$-maps are similarly combined, adding the rms values in quadrature and recalculating $z$ with the new minimum composite rms. The $z$-map from ionization equilibrium is considered separately, as a third independent constraint. Then the three ``master maps'' are added together, so that the photometric, ionization, and single-species synthesis constraints are each weighted equally in deriving the composite solution for $\Teff$ and $\logg$. For stars where one or more of these maps cannot be calculated (no photometric colors in the literature, or too few metal lines to provide meaningful synthesis constraints), the other maps usually provide sufficient information to reach a solution for $\Teff$ and $\logg$.

Because many of the parameters that we are trying to determine are strongly interdependent, we must iterate until the derived values converge --- solving first for $\vr$ and $\vsini$, then determining $\Teff$ and $\logg$, and lastly calculating $\xi$ and $\logeps$ for all species present, before recomputing the line list, re-fitting the continuum, and starting the process again. In practice, three or four iterations are usually required for all photospheric parameters to stabilize to within 5\%, at which point we declare the analysis to be complete.


\section{Photospheric analysis results and stellar classification}

Table~\ref{results} presents the final photometric parameter solutions for our target stars, based on the photometric and spectroscopic analysis described in the previous section. Only the formal random errors from the analysis are quoted for each parameter, and no attempt has been made to quantify the systematic errors in $\Teff$, $\logg$, or $\xi$, or to determine the additional uncertainty in the chemical abundances due to such errors in temperature, gravity, or microturbulence. In those cases where we could not determine $\xi$ (too few lines, or lines too broad), we set $\xi = 2.0 \pm 1.0 \kms$. The values for [Fe/H] and [Mg/H] (computed using the solar $\logeps$ values of \cite{grevesse98}) are listed primarily to aid in the identification of each star's evolutionary status, and should not be considered to be competitive with careful and detailed abundance analyses such as those reported by \cite{kinman00}. The solutions for the hot and fast-rotating stars should be regarded with particular skepticism, as the spectral continuum level is poorly defined across broad absorption lines. For one hot star, PG1530+212, we were unable to determine $\Teff$ and $\logg$ with any available techniques, so we adopted $\Teff = 15000 \K$ and $\logg = 4.0$ in order to proceed with spectral line fitting and $\vsini$ measurement. 
\clearpage
\begin{deluxetable}{lcccccccl}
\rotate
\tablewidth{0pt}
\tablecaption{Parameters derived from spectroscopic and photometric analysis. \label{results}}
\tabletypesize{\scriptsize}
\tablehead{
 	& 	& 	&$\xi$	&$\vsini$	& 	&  	&heliocentric	& 	\\
star	&$\Teff$~(K)	&$\logg$~(cgs)	&(km/s)	&(km/s)	&[Fe/H]	&[Mg/H]	&$\vr$ (km/s)	&stellar~type}
\startdata
BD$-$12~2669    	&${\phn}6880\;^{+609{\phn}}_{-510{\phn}}$    	&$3.91\;^{+0.73}_{-0.76}$    	&$1.9\;^{+1.1}_{-1.0}$    	&${\phn}32.0\;^{+3.8{\phn}}_{-3.4{\phn}}$    	&${-}2.04 \pm 0.17$    	&$< {-}0.85$    	&${\phn}{+}50.79 \pm 6.29{\phn}$    	&main~seq.    	\\
BD$-$07~0230    	&${\phn}9647\;^{+285{\phn}}_{-252{\phn}}$    	&$3.40\;^{+0.34}_{-0.31}$    	&$2.2\;^{+0.9}_{-0.9}$    	&${\phn}{\phn}2.4\;^{+3.1{\phn}}_{-2.4{\phn}}$    	&${-}0.44 \pm 0.11$    	&${-}0.14 \pm 0.11$    	&${\phn}{\phn}{-}1.67 \pm 0.80{\phn}$    	&poss.~HB    	\\
BD$-$02~0524    	&$16563\;^{+392{\phn}}_{-720{\phn}}$    	&$3.75\;^{+0.53}_{-0.68}$    	&$0.8\;^{+5.3}_{-0.8}$    	&${\phn}11.9\;^{+3.4{\phn}}_{-3.1{\phn}}$    	&${-}0.63 \pm 0.33$    	&${-}0.06 \pm 0.17$    	&${\phn}{\phn}{-}6.28 \pm 1.92{\phn}$    	&main~seq.    	\\
BD$+$00~0145    	&${\phn}9121\;^{+395{\phn}}_{-562{\phn}}$    	&$4.18\;^{+0.14}_{-0.16}$    	&$0.0\;^{+5.4}_{-0.0}$    	&${\phn}27.8\;^{+3.0{\phn}}_{-3.3{\phn}}$    	&${-}2.47 \pm 0.12$    	&${-}2.11 \pm 0.04$    	&${-}265.68 \pm 1.05{\phn}$    	&main~seq.    	\\
BD$+$01~0513    	&${\phn}7211\;^{+229{\phn}}_{-146{\phn}}$    	&$5.28\;^{+0.33}_{-0.29}$    	&$0.7\;^{+0.4}_{-0.7}$    	&${\phn}17.1\;^{+1.0{\phn}}_{-1.1{\phn}}$    	&${+}0.19 \pm 0.09$    	&$< {-}0.11$    	&${\phn}{+}13.60 \pm 1.45{\phn}$    	&subdwarf?    	\\
BD$+$01~0514    	&${\phn}7673\;^{+453{\phn}}_{-292{\phn}}$    	&$3.10\;^{+0.32}_{-0.32}$    	&$2.0\;^{+1.0}_{-1.0}$    	&$137.9\;^{+14.6}_{-14.7}$    	&${-}2.00 \pm 0.51$    	&${+}0.13 \pm 0.13$    	&${\phn}{+}11.70 \pm 19.77$    	&RR~Lyr ?    	\\
BD$+$01~0548    	&${\phn}8714\;^{+235{\phn}}_{-160{\phn}}$    	&$3.38\;^{+0.20}_{-0.12}$    	&$0.7\;^{+0.7}_{-0.7}$    	&${\phn}10.2\;^{+0.7{\phn}}_{-0.8{\phn}}$    	&${-}2.23 \pm 0.06$    	&${-}1.83 \pm 0.02$    	&${\phn}{-}55.84 \pm 0.79{\phn}$    	&HB    	\\
BD$+$03~0740    	&${\phn}6406\;^{+538{\phn}}_{-332{\phn}}$    	&$3.76\;^{+0.73}_{-0.58}$    	&$1.5\;^{+1.4}_{-1.5}$    	&${\phn}{\phn}6.0\;^{+3.5{\phn}}_{-6.0{\phn}}$    	&${-}2.87 \pm 0.46$    	&${-}2.61 \pm 0.27$    	&${+}174.86 \pm 1.52{\phn}$    	&subgiant    	\\
BD$+$09~3223    	&${\phn}5305\;^{+183{\phn}}_{-107{\phn}}$    	&$1.91\;^{+0.35}_{-0.27}$    	&$2.1\;^{+0.2}_{-0.2}$    	&${\phn}{\phn}5.4\;^{+0.6{\phn}}_{-0.9{\phn}}$    	&${-}2.34 \pm 0.10$    	&${-}1.78 \pm 0.15$    	&${\phn}{+}67.25 \pm 0.22{\phn}$    	&HB    	\\
BD$+$10~2495    	&${\phn}5275\;^{+146{\phn}}_{-102{\phn}}$    	&$2.75\;^{+0.29}_{-0.30}$    	&$2.1\;^{+0.3}_{-0.3}$    	&${\phn}{\phn}2.6\;^{+1.2{\phn}}_{-1.5{\phn}}$    	&${-}2.07 \pm 0.12$    	&${-}1.54 \pm 0.12$    	&${+}262.57 \pm 0.10{\phn}$    	&RGB    	\\
BD$+$11~2998    	&${\phn}5647\;^{+107{\phn}}_{-90{\phn}{\phn}}$    	&$2.39\;^{+0.20}_{-0.16}$    	&$2.2\;^{+0.1}_{-0.1}$    	&${\phn}{\phn}6.6\;^{+0.4{\phn}}_{-0.5{\phn}}$    	&${-}1.28 \pm 0.06$    	&${-}0.97 \pm 0.07$    	&${\phn}{+}50.54 \pm 0.20{\phn}$    	&HB    	\\
BD$+$14~4757    	&${\phn}6390\;^{+133{\phn}}_{-106{\phn}}$    	&$4.99\;^{+0.21}_{-0.26}$    	&$1.2\;^{+0.3}_{-0.3}$    	&${\phn}{\phn}2.1\;^{+0.8{\phn}}_{-1.2{\phn}}$    	&${-}0.56 \pm 0.10$    	&${-}0.68 \pm 0.10$    	&${\phn}{+}25.37 \pm 0.19{\phn}$    	&main~seq.    	\\
BD$+$17~3248    	&${\phn}5398\;^{+221{\phn}}_{-83{\phn}{\phn}}$    	&$2.21\;^{+0.37}_{-0.20}$    	&$2.1\;^{+0.2}_{-0.2}$    	&${\phn}{\phn}5.4\;^{+0.8{\phn}}_{-1.0{\phn}}$    	&${-}2.08 \pm 0.07$    	&${-}1.68 \pm 0.11$    	&${-}145.23 \pm 0.87{\phn}$    	&HB    	\\
BD$+$17~4708    	&${\phn}6297\;^{+104{\phn}}_{-147{\phn}}$    	&$4.40\;^{+0.21}_{-0.31}$    	&$1.5\;^{+0.2}_{-0.2}$    	&${\phn}{\phn}3.5\;^{+0.6{\phn}}_{-0.7{\phn}}$    	&${-}1.61 \pm 0.05$    	&${-}1.15 \pm 0.08$    	&${-}287.52 \pm 0.42{\phn}$    	&main~seq.    	\\
BD$+$18~0153    	&${\phn}6554\;^{+143{\phn}}_{-85{\phn}{\phn}}$    	&$4.16\;^{+0.17}_{-0.17}$    	&$2.0\;^{+1.0}_{-1.0}$    	&${\phn}81.3\;^{+4.8{\phn}}_{-4.7{\phn}}$    	&${+}0.88 \pm 0.21$    	&${+}0.15 \pm 0.19$    	&${\phn}{\phn}{-}2.79 \pm 3.46{\phn}$    	&main~seq.    	\\
BD$+$18~2757    	&${\phn}4741\;^{+152{\phn}}_{-72{\phn}{\phn}}$    	&$1.16\;^{+0.18}_{-0.16}$    	&$1.3\;^{+1.0}_{-1.0}$    	&${\phn}{\phn}5.5\;^{+1.0{\phn}}_{-0.9{\phn}}$    	&${-}2.43 \pm 0.12$    	&${-}1.75 \pm 0.16$    	&${\phn}{-}24.92 \pm 0.30{\phn}$    	&RGB    	\\
BD$+$18~2890    	&${\phn}5347\;^{+123{\phn}}_{-125{\phn}}$    	&$2.60\;^{+0.24}_{-0.35}$    	&$1.9\;^{+0.2}_{-0.2}$    	&${\phn}{\phn}3.2\;^{+0.9{\phn}}_{-1.6{\phn}}$    	&${-}1.78 \pm 0.15$    	&${-}0.83 \pm 0.08$    	&${\phn}{-}31.05 \pm 0.19{\phn}$    	&HB    	\\
BD$+$20~3004    	&$14549\;^{+268{\phn}}_{-627{\phn}}$    	&$3.80\;^{+0.56}_{-0.81}$    	&$2.0\;^{+1.0}_{-1.0}$    	&$104.7\;^{+19.8}_{-15.4}$    	&${-}0.91 \pm 0.83$    	&${-}0.17 \pm 0.78$    	&${\phn}{+}21.68 \pm 14.28$    	&main~seq.    	\\
BD$+$25~1981    	&${\phn}7302\;^{+363{\phn}}_{-312{\phn}}$    	&$4.41\;^{+0.34}_{-0.44}$    	&$2.1\;^{+0.6}_{-0.5}$    	&${\phn}{\phn}7.9\;^{+1.4{\phn}}_{-1.3{\phn}}$    	&${-}1.43 \pm 0.13$    	&${-}1.18 \pm 0.13$    	&${\phn}{+}59.60 \pm 1.02{\phn}$    	&main~seq.    	\\
BD$+$25~2436    	&${\phn}4847\;^{+38{\phn}{\phn}}_{-44{\phn}{\phn}}$    	&$2.14\;^{+0.35}_{-0.29}$    	&$1.7\;^{+0.2}_{-0.2}$    	&${\phn}{\phn}6.1\;^{+0.6{\phn}}_{-0.6{\phn}}$    	&${-}0.76 \pm 0.15$    	&${-}0.22 \pm 0.19$    	&${\phn}{+}24.27 \pm 0.26{\phn}$    	&RGB    	\\
BD$+$25~2459    	&${\phn}4743\;^{+112{\phn}}_{-152{\phn}}$    	&$2.78\;^{+0.43}_{-0.66}$    	&$1.5\;^{+0.2}_{-0.2}$    	&${\phn}{\phn}4.9\;^{+0.7{\phn}}_{-0.9{\phn}}$    	&${-}0.28 \pm 0.16$    	&${-}0.33 \pm 0.22$    	&${\phn}{+}21.16 \pm 0.23{\phn}$    	&RGB    	\\
BD$+$25~2497    	&${\phn}5169\;^{+62{\phn}{\phn}}_{-71{\phn}{\phn}}$    	&$2.42\;^{+0.22}_{-0.04}$    	&$1.3\;^{+0.2}_{-0.2}$    	&${\phn}{\phn}5.1\;^{+0.7{\phn}}_{-0.7{\phn}}$    	&${-}0.84 \pm 0.20$    	&${+}0.01 \pm 0.11$    	&${\phn}{+}76.09 \pm 0.21{\phn}$    	&RGB    	\\
BD$+$25~2602    	&${\phn}8250\;^{+502{\phn}}_{-408{\phn}}$    	&$3.26\;^{+0.23}_{-0.30}$    	&$2.3\;^{+1.5}_{-0.9}$    	&${\phn}13.3\;^{+1.7{\phn}}_{-1.8{\phn}}$    	&${-}2.08 \pm 0.11$    	&${-}1.60 \pm 0.08$    	&${\phn}{-}66.23 \pm 1.54{\phn}$    	&HB    	\\
BD$+$27~2057    	&${\phn}4695\;^{+47{\phn}{\phn}}_{-45{\phn}{\phn}}$    	&$1.58\;^{+0.24}_{-0.28}$    	&$1.7\;^{+0.2}_{-0.2}$    	&${\phn}{\phn}6.5\;^{+0.7{\phn}}_{-0.7{\phn}}$    	&${-}1.25 \pm 0.15$    	&${-}0.26 \pm 0.28$    	&${\phn}{-}33.10 \pm 0.28{\phn}$    	&RGB    	\\
BD$+$29~2231    	&${\phn}4756\;^{+54{\phn}{\phn}}_{-51{\phn}{\phn}}$    	&$2.39\;^{+0.25}_{-0.38}$    	&$1.7\;^{+0.2}_{-0.2}$    	&${\phn}{\phn}4.8\;^{+0.8{\phn}}_{-1.1{\phn}}$    	&${-}0.65 \pm 0.15$    	&${-}0.48 \pm 0.25$    	&${\phn}{+}22.31 \pm 0.25{\phn}$    	&RGB    	\\
BD$+$29~2294    	&${\phn}5132\;^{+68{\phn}{\phn}}_{-51{\phn}{\phn}}$    	&$3.18\;^{+0.26}_{-0.44}$    	&$1.6\;^{+0.2}_{-0.2}$    	&${\phn}{\phn}5.0\;^{+0.6{\phn}}_{-0.7{\phn}}$    	&${-}0.55 \pm 0.14$    	&${-}0.42 \pm 0.07$    	&${\phn}{-}14.52 \pm 0.20{\phn}$    	&RGB    	\\
BD$+$30~2338    	&${\phn}7675\;^{+298{\phn}}_{-283{\phn}}$    	&$4.90\;^{+0.19}_{-0.35}$    	&$1.6\;^{+0.5}_{-0.5}$    	&${\phn}16.8\;^{+1.4{\phn}}_{-1.1{\phn}}$    	&${+}0.25 \pm 0.12$    	&${+}0.06 \pm 0.11$    	&${\phn}{\phn}{+}6.72 \pm 0.71{\phn}$    	&main~seq.    	\\
BD$+$30~2355    	&$10215\;^{+132{\phn}}_{-158{\phn}}$    	&$3.29\;^{+0.34}_{-0.57}$    	&$2.0\;^{+1.0}_{-1.0}$    	&${\phn}97.3\;^{+21.4}_{-27.7}$    	&${-}2.70 \pm 0.35$    	&\nodata    	&${\phn}{\phn}{+}1.20 \pm 24.98$    	&poss.~HB    	\\
BD$+$30~2431    	&$16111\;^{+413{\phn}}_{-581{\phn}}$    	&$3.78\;^{+0.65}_{-0.63}$    	&$0.0\;^{+1.5}_{-0.0}$    	&${\phn}{\phn}0.0\;^{+4.2{\phn}}_{-0.0{\phn}}$    	&${+}0.38 \pm 0.19$    	&$< {-}2.00$    	&${\phn}{-}26.54 \pm 0.80{\phn}$    	&HB    	\\
BD$+$32~2188    	&$10257\;^{+233{\phn}}_{-218{\phn}}$    	&$2.00\;^{+0.26}_{-0.23}$    	&$0.7\;^{+0.4}_{-0.7}$    	&${\phn}{\phn}0.4\;^{+0.8{\phn}}_{-0.4{\phn}}$    	&${-}1.05 \pm 0.06$    	&${-}0.99 \pm 0.07$    	&${\phn}{+}92.82 \pm 0.23{\phn}$    	&post-AGB    	\\
BD$+$33~2171    	&${\phn}7149\;^{+199{\phn}}_{-140{\phn}}$    	&$3.72\;^{+0.17}_{-0.33}$    	&$3.0\;^{+0.3}_{-0.3}$    	&${\phn}44.0\;^{+1.1{\phn}}_{-1.1{\phn}}$    	&${-}1.73 \pm 0.07$    	&${-}1.23 \pm 0.03$    	&${\phn}{+}53.05 \pm 0.71{\phn}$    	&RR~Lyr ?    	\\
BD$+$33~2642    	&$16321\;^{+2719}_{-1136}$    	&$1.99\;^{+1.00}_{-0.64}$    	&$2.0\;^{+1.0}_{-1.0}$    	&${\phn}19.3\;^{+4.4{\phn}}_{-5.0{\phn}}$    	&$< 0.17$    	&$< {-}0.23$    	&${\phn}{-}94.71 \pm 2.50{\phn}$    	&post-AGB    	\\
BD$+$34~2371    	&${\phn}5005\;^{+49{\phn}{\phn}}_{-63{\phn}{\phn}}$    	&$2.53\;^{+0.28}_{-0.34}$    	&$0.8\;^{+0.3}_{-0.4}$    	&${\phn}{\phn}4.5\;^{+0.8{\phn}}_{-0.8{\phn}}$    	&${-}0.41 \pm 0.20$    	&${-}0.11 \pm 0.25$    	&${\phn}{-}18.96 \pm 0.22{\phn}$    	&RGB    	\\
BD$+$36~2242    	&$11650\;^{+118{\phn}}_{-174{\phn}}$    	&$4.01\;^{+0.18}_{-0.39}$    	&$2.0\;^{+1.0}_{-1.0}$    	&${\phn}77.1\;^{+3.1{\phn}}_{-5.3{\phn}}$    	&${-}0.84 \pm 0.38$    	&${-}0.86 \pm 0.12$    	&${\phn}{\phn}{+}2.45 \pm 4.08{\phn}$    	&main~seq.    	\\
BD$+$36~2268    	&$19832\;^{+820{\phn}}_{-1996}$    	&$3.82\;^{+0.98}_{-1.01}$    	&$0.0\;^{+9.0}_{-0.0}$    	&${\phn}51.7\;^{+8.7{\phn}}_{-7.8{\phn}}$    	&${-}0.26 \pm 5.61$    	&${-}0.13 \pm 0.53$    	&${\phn}{+}40.97 \pm 5.93{\phn}$    	&main~seq.    	\\
BD$+$36~2303    	&${\phn}4705\;^{+30{\phn}{\phn}}_{-20{\phn}{\phn}}$    	&$2.40\;^{+0.14}_{-0.11}$    	&$1.7\;^{+0.2}_{-0.2}$    	&${\phn}{\phn}5.1\;^{+0.7{\phn}}_{-0.8{\phn}}$    	&${-}0.77 \pm 0.19$    	&${-}0.63 \pm 0.13$    	&${\phn}{+}80.47 \pm 0.22{\phn}$    	&RGB    	\\
BD$+$42~2309    	&${\phn}8796\;^{+186{\phn}}_{-223{\phn}}$    	&$3.39\;^{+0.15}_{-0.15}$    	&$2.3\;^{+0.3}_{-0.3}$    	&${\phn}30.7\;^{+0.9{\phn}}_{-1.1{\phn}}$    	&${-}1.69 \pm 0.05$    	&${-}1.33 \pm 0.02$    	&${-}142.34 \pm 0.70{\phn}$    	&HB    	\\
BD$+$46~1998    	&${\phn}6811\;^{+163{\phn}}_{-140{\phn}}$    	&$4.31\;^{+0.35}_{-0.28}$    	&$1.2\;^{+0.6}_{-0.8}$    	&${\phn}31.8\;^{+1.5{\phn}}_{-1.6{\phn}}$    	&${+}0.10 \pm 0.12$    	&$< {-}0.65$    	&${\phn}{-}11.74 \pm 1.30{\phn}$    	&main~seq.    	\\
BD$+$49~2137    	&$15047\;^{+334{\phn}}_{-432{\phn}}$    	&$3.90\;^{+0.34}_{-0.72}$    	&$4.9\;^{+3.1}_{-3.1}$    	&${\phn}32.2\;^{+8.8{\phn}}_{-8.0{\phn}}$    	&${+}0.28 \pm 0.37$    	&${-}0.19 \pm 0.52$    	&${\phn}{+}93.54 \pm 4.60{\phn}$    	&poss. HB    	\\
BPS~CS~22189-5    	&${\phn}7397\;^{+219{\phn}}_{-239{\phn}}$    	&$3.60\;^{+0.41}_{-0.37}$    	&$3.0\;^{+0.3}_{-0.3}$    	&${\phn}13.9\;^{+0.8{\phn}}_{-0.8{\phn}}$    	&${-}1.00 \pm 0.08$    	&${-}1.26 \pm 0.08$    	&${-}107.48 \pm 0.46{\phn}$    	&unknown    	\\
BPS~CS~22894-36    	&${\phn}7832\;^{+1358}_{-673{\phn}}$    	&$4.12\;^{+1.47}_{-1.02}$    	&$3.3\;^{+2.1}_{-1.4}$    	&${\phn}10.2\;^{+3.3{\phn}}_{-2.8{\phn}}$    	&${-}1.77 \pm 0.20$    	&${-}2.28 \pm 0.26$    	&${-}267.70 \pm 1.39{\phn}$    	&main~seq.    	\\
Feige~40    	&$15904\;^{+696{\phn}}_{-693{\phn}}$    	&$4.54\;^{+0.52}_{-0.76}$    	&$2.0\;^{+1.0}_{-1.0}$    	&$120.7\;^{+14.5}_{-13.3}$    	&${-}1.71 \pm 0.56$    	&${-}0.58 \pm 0.85$    	&${\phn}{+}74.19 \pm 14.07$    	&main~seq.    	\\
Feige~41    	&$10023\;^{+308{\phn}}_{-409{\phn}}$    	&$4.42\;^{+0.20}_{-0.19}$    	&$1.8\;^{+1.5}_{-1.2}$    	&${\phn}{\phn}2.7\;^{+2.4{\phn}}_{-2.7{\phn}}$    	&${-}0.33 \pm 0.09$    	&${-}0.32 \pm 0.10$    	&${\phn}{-}23.90 \pm 0.55{\phn}$    	&main~seq.    	\\
Feige~84    	&$18587\;^{+748{\phn}}_{-969{\phn}}$    	&$4.45\;^{+0.59}_{-1.03}$    	&$2.0\;^{+1.0}_{-1.0}$    	&$111.0\;^{+31.7}_{-30.5}$    	&$< {-}1.54$    	&$< {-}0.12$    	&${+}153.04 \pm 27.76$    	&main~seq.    	\\
GCRV~63536    	&${\phn}8702\;^{+154{\phn}}_{-119{\phn}}$    	&$4.29\;^{+0.21}_{-0.21}$    	&$3.4\;^{+0.2}_{-0.2}$    	&${\phn}10.9\;^{+0.6{\phn}}_{-0.4{\phn}}$    	&${-}0.83 \pm 0.05$    	&${-}1.10 \pm 0.05$    	&${\phn}{+}27.05 \pm 0.24{\phn}$    	&main~seq.    	\\
HD~97    	&${\phn}5270\;^{+125{\phn}}_{-111{\phn}}$    	&$2.83\;^{+0.33}_{-0.24}$    	&$1.2\;^{+0.3}_{-0.3}$    	&${\phn}{\phn}4.0\;^{+1.0{\phn}}_{-1.3{\phn}}$    	&${-}1.42 \pm 0.12$    	&${-}1.19 \pm 0.15$    	&${\phn}{+}75.72 \pm 0.46{\phn}$    	&RGB    	\\
HD~1112    	&$11352\;^{+129{\phn}}_{-222{\phn}}$    	&$4.04\;^{+0.24}_{-0.44}$    	&$2.0\;^{+1.0}_{-1.0}$    	&$164.9\;^{+11.8}_{-11.4}$    	&\nodata    	&${-}0.25 \pm 0.11$    	&${\phn}{\phn}{-}9.41 \pm 9.70{\phn}$    	&main~seq.    	\\
HD~2857    	&${\phn}8002\;^{+444{\phn}}_{-371{\phn}}$    	&$3.38\;^{+0.29}_{-0.35}$    	&$4.5\;^{+2.1}_{-1.2}$    	&${\phn}25.1\;^{+2.5{\phn}}_{-2.4{\phn}}$    	&${-}1.67 \pm 0.11$    	&${-}1.54 \pm 0.07$    	&${-}155.25 \pm 1.95{\phn}$    	&HB    	\\
HD~2880    	&${\phn}4810\;^{+92{\phn}{\phn}}_{-67{\phn}{\phn}}$    	&$4.01\;^{+0.59}_{-0.41}$    	&$2.0\;^{+1.0}_{-1.0}$    	&${\phn}{\phn}6.5\;^{+1.2{\phn}}_{-1.4{\phn}}$    	&${-}0.83 \pm 0.17$    	&${-}0.07 \pm 0.68$    	&${\phn}{-}10.56 \pm 0.75{\phn}$    	&main~seq.    	\\
HD~6229    	&${\phn}5200\;^{+185{\phn}}_{-103{\phn}}$    	&$1.84\;^{+0.35}_{-0.25}$    	&$1.6\;^{+0.3}_{-0.3}$    	&${\phn}{\phn}5.7\;^{+1.0{\phn}}_{-1.0{\phn}}$    	&${-}1.35 \pm 0.14$    	&${-}0.45 \pm 0.17$    	&${\phn}{-}89.13 \pm 0.42{\phn}$    	&HB    	\\
HD~6461    	&${\phn}5109\;^{+53{\phn}{\phn}}_{-61{\phn}{\phn}}$    	&$1.86\;^{+0.30}_{-0.36}$    	&$1.6\;^{+0.3}_{-0.3}$    	&${\phn}{\phn}6.2\;^{+0.8{\phn}}_{-0.8{\phn}}$    	&${-}1.30 \pm 0.15$    	&${-}0.32 \pm 0.21$    	&${\phn}{\phn}{+}7.99 \pm 0.64{\phn}$    	&HB    	\\
HD~7374    	&$13327\;^{+162{\phn}}_{-193{\phn}}$    	&$3.84\;^{+0.32}_{-0.63}$    	&$4.0\;^{+1.0}_{-1.0}$    	&${\phn}20.8\;^{+1.5{\phn}}_{-1.6{\phn}}$    	&${-}0.80 \pm 0.08$    	&${-}0.83 \pm 0.09$    	&${\phn}{-}14.48 \pm 0.81{\phn}$    	&main seq. CP    	\\
HD~8376    	&${\phn}7606\;^{+552{\phn}}_{-275{\phn}}$    	&$2.87\;^{+0.46}_{-0.37}$    	&$3.3\;^{+3.1}_{-3.3}$    	&${\phn}{\phn}0.0\;^{+6.8{\phn}}_{-0.0{\phn}}$    	&${-}3.06 \pm 0.13$    	&${-}2.57 \pm 0.08$    	&${+}146.76 \pm 1.20{\phn}$    	&HB    	\\
HD~13978    	&${\phn}7060\;^{+876{\phn}}_{-422{\phn}}$    	&$5.12\;^{+0.83}_{-1.40}$    	&$2.0\;^{+1.0}_{-1.0}$    	&${\phn}91.8\;^{+6.0{\phn}}_{-9.4{\phn}}$    	&${-}0.40 \pm 0.40$    	&${+}0.53 \pm 0.13$    	&${\phn}{+}18.22 \pm 5.74{\phn}$    	&subdwarf?    	\\
HD~14829    	&${\phn}9086\;^{+267{\phn}}_{-222{\phn}}$    	&$3.31\;^{+0.15}_{-0.22}$    	&$0.0\;^{+8.0}_{-0.0}$    	&${\phn}14.3\;^{+5.6{\phn}}_{-5.3{\phn}}$    	&${-}2.01 \pm 0.31$    	&${-}1.87 \pm 0.08$    	&${-}173.22 \pm 3.28{\phn}$    	&HB    	\\
HD~24000    	&${\phn}9439\;^{+441{\phn}}_{-616{\phn}}$    	&$3.91\;^{+0.15}_{-0.21}$    	&$2.0\;^{+1.0}_{-1.0}$    	&${\phn}58.2\;^{+11.2}_{-9.6{\phn}}$    	&${+}0.03 \pm 0.52$    	&${-}0.05 \pm 0.20$    	&${\phn}{+}27.12 \pm 5.83{\phn}$    	&main~seq.    	\\
HD~24341    	&${\phn}5348\;^{+118{\phn}}_{-117{\phn}}$    	&$3.79\;^{+0.27}_{-0.37}$    	&$1.2\;^{+0.3}_{-0.3}$    	&${\phn}{\phn}3.5\;^{+1.3{\phn}}_{-1.9{\phn}}$    	&${-}0.90 \pm 0.09$    	&${-}0.37 \pm 0.09$    	&${+}144.34 \pm 0.34{\phn}$    	&subgiant    	\\
HD~25532    	&${\phn}5553\;^{+83{\phn}{\phn}}_{-78{\phn}{\phn}}$    	&$2.11\;^{+0.16}_{-0.16}$    	&$2.5\;^{+0.1}_{-0.1}$    	&${\phn}{\phn}7.7\;^{+0.5{\phn}}_{-0.3{\phn}}$    	&${-}1.41 \pm 0.06$    	&${-}0.71 \pm 0.09$    	&${-}110.39 \pm 0.32{\phn}$    	&HB    	\\
HD~27295    	&$11956\;^{+155{\phn}}_{-228{\phn}}$    	&$3.92\;^{+0.17}_{-0.26}$    	&$1.4\;^{+0.7}_{-1.0}$    	&${\phn}{\phn}4.0\;^{+1.3{\phn}}_{-1.5{\phn}}$    	&${-}0.95 \pm 0.06$    	&${-}0.46 \pm 0.05$    	&${\phn}{\phn}{+}5.84 \pm 0.37{\phn}$    	&main seq. CP    	\\
HD~60778    	&${\phn}8020\;^{+218{\phn}}_{-199{\phn}}$    	&$3.13\;^{+0.19}_{-0.18}$    	&$2.6\;^{+0.5}_{-0.4}$    	&${\phn}10.0\;^{+0.9{\phn}}_{-1.0{\phn}}$    	&${-}1.48 \pm 0.06$    	&${-}0.99 \pm 0.10$    	&${\phn}{+}42.58 \pm 1.06{\phn}$    	&HB    	\\
HD~63791    	&${\phn}4954\;^{+105{\phn}}_{-56{\phn}{\phn}}$    	&$2.17\;^{+0.20}_{-0.20}$    	&$1.6\;^{+0.2}_{-0.2}$    	&${\phn}{\phn}4.4\;^{+0.8{\phn}}_{-0.9{\phn}}$    	&${-}1.72 \pm 0.07$    	&${-}1.20 \pm 0.10$    	&${-}106.97 \pm 0.73{\phn}$    	&RGB    	\\
HD~64488    	&${\phn}8826\;^{+307{\phn}}_{-378{\phn}}$    	&$3.63\;^{+0.11}_{-0.12}$    	&$2.0\;^{+2.0}_{-2.0}$    	&$150.6\;^{+12.8}_{-17.3}$    	&${-}0.77 \pm 0.23$    	&${-}0.02 \pm 0.13$    	&${\phn}{\phn}{+}5.08 \pm 8.74{\phn}$    	&main~seq.    	\\
HD~74721    	&${\phn}8677\;^{+210{\phn}}_{-140{\phn}}$    	&$3.38\;^{+0.15}_{-0.10}$    	&$2.1\;^{+0.8}_{-0.6}$    	&${\phn}{\phn}2.6\;^{+1.4{\phn}}_{-2.6{\phn}}$    	&${-}1.41 \pm 0.04$    	&${-}0.97 \pm 0.05$    	&${\phn}{+}31.41 \pm 0.36{\phn}$    	&HB    	\\
HD~76431    	&$29509\;^{+1087}_{-1238}$    	&$4.18\;^{+0.44}_{-0.41}$    	&$1.6\;^{+1.8}_{-1.6}$    	&${\phn}{\phn}0.0\;^{+3.5{\phn}}_{-0.0{\phn}}$    	&\nodata    	&${-}0.66 \pm 0.05$    	&${\phn}{+}46.91 \pm 0.73{\phn}$    	&main~seq.    	\\
HD~79452    	&${\phn}5042\;^{+60{\phn}{\phn}}_{-58{\phn}{\phn}}$    	&$2.14\;^{+0.16}_{-0.22}$    	&$1.4\;^{+0.1}_{-0.2}$    	&${\phn}{\phn}6.1\;^{+0.5{\phn}}_{-0.4{\phn}}$    	&${-}0.91 \pm 0.10$    	&${-}0.12 \pm 0.09$    	&${\phn}{+}55.77 \pm 0.49{\phn}$    	&HB    	\\
HD~79530    	&${\phn}7224\;^{+178{\phn}}_{-169{\phn}}$    	&$5.00\;^{+0.34}_{-0.24}$    	&$1.5\;^{+0.4}_{-0.4}$    	&${\phn}19.9\;^{+1.0{\phn}}_{-0.9{\phn}}$    	&${-}0.02 \pm 0.14$    	&${-}0.12 \pm 0.10$    	&${\phn}{+}21.08 \pm 0.91{\phn}$    	&subdwarf?    	\\
HD~82590    	&${\phn}6094\;^{+492{\phn}}_{-272{\phn}}$    	&$2.04\;^{+0.81}_{-0.52}$    	&$2.3\;^{+0.5}_{-0.4}$    	&${\phn}{\phn}9.7\;^{+1.1{\phn}}_{-1.0{\phn}}$    	&${-}1.50 \pm 0.10$    	&${-}1.02 \pm 0.11$    	&${+}214.69 \pm 0.49{\phn}$    	&RR~Lyr ?    	\\
HD~83751    	&${\phn}9735\;^{+119{\phn}}_{-170{\phn}}$    	&$3.32\;^{+0.22}_{-0.33}$    	&$2.3\;^{+0.5}_{-0.6}$    	&${\phn}25.5\;^{+0.8{\phn}}_{-0.8{\phn}}$    	&${-}0.22 \pm 0.06$    	&${-}0.09 \pm 0.09$    	&${\phn}{+}11.93 \pm 0.97{\phn}$    	&poss.~HB    	\\
HD~86986    	&${\phn}7775\;^{+237{\phn}}_{-231{\phn}}$    	&$3.05\;^{+0.21}_{-0.24}$    	&$2.5\;^{+0.6}_{-0.5}$    	&${\phn}{\phn}9.2\;^{+0.9{\phn}}_{-0.8{\phn}}$    	&${-}1.85 \pm 0.05$    	&${-}1.35 \pm 0.06$    	&${\phn}{+}10.13 \pm 0.50{\phn}$    	&HB    	\\
HD~87047    	&${\phn}7682\;^{+397{\phn}}_{-302{\phn}}$    	&$2.95\;^{+0.35}_{-0.39}$    	&$1.2\;^{+1.3}_{-1.2}$    	&${\phn}{\phn}9.2\;^{+1.4{\phn}}_{-1.5{\phn}}$    	&${-}2.36 \pm 0.10$    	&${-}1.93 \pm 0.06$    	&${+}138.52 \pm 0.80{\phn}$    	&HB    	\\
HD~87112    	&${\phn}9557\;^{+224{\phn}}_{-268{\phn}}$    	&$3.46\;^{+0.23}_{-0.23}$    	&$1.9\;^{+1.4}_{-1.4}$    	&${\phn}{\phn}7.2\;^{+1.9{\phn}}_{-1.6{\phn}}$    	&${-}1.65 \pm 0.07$    	&${-}1.13 \pm 0.06$    	&${-}172.55 \pm 0.69{\phn}$    	&HB    	\\
HD~93329    	&${\phn}8042\;^{+415{\phn}}_{-299{\phn}}$    	&$3.09\;^{+0.22}_{-0.26}$    	&$2.7\;^{+0.8}_{-0.6}$    	&${\phn}{\phn}9.6\;^{+1.3{\phn}}_{-1.3{\phn}}$    	&${-}1.49 \pm 0.08$    	&${-}1.07 \pm 0.12$    	&${+}204.97 \pm 0.71{\phn}$    	&HB    	\\
HD~97560    	&${\phn}5422\;^{+134{\phn}}_{-42{\phn}{\phn}}$    	&$2.39\;^{+0.25}_{-0.24}$    	&$1.9\;^{+0.2}_{-0.2}$    	&${\phn}{\phn}7.4\;^{+0.3{\phn}}_{-0.6{\phn}}$    	&${-}1.06 \pm 0.08$    	&${-}0.73 \pm 0.13$    	&${\phn}{-}20.16 \pm 0.35{\phn}$    	&HB    	\\
HD~100340    	&$24005\;^{+781{\phn}}_{-1402}$    	&$4.12\;^{+0.68}_{-0.73}$    	&$2.0\;^{+1.0}_{-1.0}$    	&$156.0\;^{+14.6}_{-13.1}$    	&\nodata    	&${+}0.14 \pm 0.49$    	&${+}256.89 \pm 19.94$    	&main~seq.    	\\
HD~103376    	&$13554\;^{+267{\phn}}_{-326{\phn}}$    	&$3.96\;^{+0.60}_{-0.82}$    	&$2.0\;^{+1.0}_{-1.0}$    	&$189.9\;^{+27.2}_{-15.8}$    	&${-}0.71 \pm 0.49$    	&${-}0.75 \pm 0.38$    	&${\phn}{\phn}{+}7.85 \pm 17.61$    	&main~seq.    	\\
HD~105183    	&$14553\;^{+190{\phn}}_{-338{\phn}}$    	&$3.63\;^{+0.38}_{-0.54}$    	&$2.0\;^{+2.0}_{-2.0}$    	&${\phn}70.2\;^{+15.0}_{-11.8}$    	&${+}0.26 \pm 0.32$    	&${-}0.19 \pm 0.24$    	&${\phn}{+}32.57 \pm 15.33$    	&main~seq.    	\\
HD~105262    	&${\phn}8855\;^{+120{\phn}}_{-62{\phn}{\phn}}$    	&$1.82\;^{+0.28}_{-0.19}$    	&$1.5\;^{+0.7}_{-0.5}$    	&${\phn}{\phn}6.1\;^{+1.1{\phn}}_{-0.7{\phn}}$    	&${-}1.61 \pm 0.04$    	&${-}1.49 \pm 0.05$    	&${\phn}{+}43.92 \pm 0.40{\phn}$    	&post-AGB    	\\
HD~105546    	&${\phn}5299\;^{+125{\phn}}_{-89{\phn}{\phn}}$    	&$2.20\;^{+0.21}_{-0.17}$    	&$2.1\;^{+0.2}_{-0.2}$    	&${\phn}{\phn}5.2\;^{+0.5{\phn}}_{-0.7{\phn}}$    	&${-}1.67 \pm 0.07$    	&${-}0.96 \pm 0.15$    	&${\phn}{+}20.94 \pm 0.72{\phn}$    	&HB    	\\
HD~105944    	&${\phn}5759\;^{+91{\phn}{\phn}}_{-63{\phn}{\phn}}$    	&$4.06\;^{+0.23}_{-0.30}$    	&$1.5\;^{+0.2}_{-0.2}$    	&${\phn}{\phn}3.6\;^{+0.8{\phn}}_{-1.2{\phn}}$    	&${-}0.27 \pm 0.13$    	&${-}0.26 \pm 0.09$    	&${\phn}{-}14.46 \pm 0.60{\phn}$    	&main~seq.    	\\
HD~108577    	&${\phn}5192\;^{+115{\phn}}_{-113{\phn}}$    	&$1.50\;^{+0.20}_{-0.20}$    	&$2.2\;^{+0.2}_{-0.2}$    	&${\phn}{\phn}5.8\;^{+0.7{\phn}}_{-0.5{\phn}}$    	&${-}2.33 \pm 0.06$    	&${-}1.88 \pm 0.29$    	&${-}111.91 \pm 0.97{\phn}$    	&AGB    	\\
HD~109995    	&${\phn}8382\;^{+436{\phn}}_{-373{\phn}}$    	&$3.25\;^{+0.16}_{-0.18}$    	&$2.7\;^{+0.7}_{-0.5}$    	&${\phn}22.9\;^{+2.0{\phn}}_{-1.7{\phn}}$    	&${-}1.76 \pm 0.09$    	&${-}1.41 \pm 0.08$    	&${-}126.31 \pm 1.87{\phn}$    	&HB    	\\
HD~110679    	&${\phn}5001\;^{+36{\phn}{\phn}}_{-31{\phn}{\phn}}$    	&$1.91\;^{+0.16}_{-0.17}$    	&$1.6\;^{+0.2}_{-0.2}$    	&${\phn}{\phn}5.1\;^{+0.6{\phn}}_{-0.7{\phn}}$    	&${-}1.08 \pm 0.18$    	&${-}0.05 \pm 0.09$    	&${\phn}{-}54.41 \pm 1.01{\phn}$    	&HB    	\\
HD~110930    	&${\phn}4934\;^{+39{\phn}{\phn}}_{-45{\phn}{\phn}}$    	&$2.31\;^{+0.55}_{-0.46}$    	&$2.0\;^{+0.2}_{-0.2}$    	&${\phn}{\phn}4.7\;^{+0.8{\phn}}_{-0.7{\phn}}$    	&${-}0.94 \pm 0.14$    	&${-}0.60 \pm 0.16$    	&${\phn}{+}67.19 \pm 0.74{\phn}$    	&RGB    	\\
HD~112030    	&${\phn}4699\;^{+100{\phn}}_{-80{\phn}{\phn}}$    	&$1.81\;^{+0.27}_{-0.24}$    	&$2.1\;^{+0.2}_{-0.2}$    	&${\phn}{\phn}4.3\;^{+0.8{\phn}}_{-1.0{\phn}}$    	&${-}1.12 \pm 0.12$    	&${-}0.69 \pm 0.19$    	&${\phn}{-}13.12 \pm 0.74{\phn}$    	&RGB    	\\
HD~112414    	&${\phn}9227\;^{+1316}_{-1245}$    	&$5.19\;^{+0.53}_{-1.03}$    	&$2.0\;^{+1.0}_{-1.0}$    	&$150.2\;^{+13.0}_{-11.6}$    	&${+}0.27 \pm 0.25$    	&${+}0.84 \pm 0.13$    	&${\phn}{-}13.18 \pm 13.58$    	&subdwarf?    	\\
HD~112693    	&${\phn}7838\;^{+198{\phn}}_{-229{\phn}}$    	&$4.68\;^{+0.09}_{-0.34}$    	&$2.0\;^{+1.0}_{-1.0}$    	&${\phn}94.5\;^{+9.7{\phn}}_{-6.4{\phn}}$    	&${+}0.26 \pm 0.20$    	&${+}0.05 \pm 0.09$    	&${\phn}{-}14.07 \pm 4.73{\phn}$    	&main~seq.    	\\
HD~112734    	&${\phn}7972\;^{+264{\phn}}_{-439{\phn}}$    	&$4.43\;^{+0.22}_{-0.44}$    	&$2.0\;^{+1.0}_{-1.0}$    	&${\phn}91.5\;^{+6.2{\phn}}_{-3.6{\phn}}$    	&${-}0.12 \pm 0.29$    	&${+}0.14 \pm 0.09$    	&${\phn}{\phn}{-}1.04 \pm 17.94$    	&main~seq.    	\\
HD~114839    	&${\phn}7618\;^{+233{\phn}}_{-315{\phn}}$    	&$4.39\;^{+0.29}_{-0.52}$    	&$2.0\;^{+1.0}_{-1.0}$    	&${\phn}66.7\;^{+5.4{\phn}}_{-3.9{\phn}}$    	&${+}0.04 \pm 0.15$    	&\nodata    	&${\phn}{-}10.67 \pm 2.41{\phn}$    	&main~seq.    	\\
HD~114930    	&${\phn}7578\;^{+237{\phn}}_{-503{\phn}}$    	&$4.25\;^{+0.36}_{-0.81}$    	&$2.0\;^{+1.0}_{-1.0}$    	&${\phn}60.9\;^{+2.9{\phn}}_{-2.5{\phn}}$    	&${-}0.13 \pm 0.17$    	&$< 0.62$    	&${\phn}{\phn}{+}4.33 \pm 3.19{\phn}$    	&main~seq.    	\\
HD~115444    	&${\phn}4736\;^{+78{\phn}{\phn}}_{-55{\phn}{\phn}}$    	&$1.62\;^{+0.19}_{-0.18}$    	&$2.2\;^{+0.3}_{-0.3}$    	&${\phn}{\phn}4.6\;^{+0.7{\phn}}_{-1.0{\phn}}$    	&${-}3.18 \pm 0.10$    	&${-}2.68 \pm 0.12$    	&${\phn}{-}26.26 \pm 0.40{\phn}$    	&RGB    	\\
HD~115520    	&${\phn}8199\;^{+449{\phn}}_{-317{\phn}}$    	&$4.63\;^{+0.34}_{-0.23}$    	&$2.0\;^{+1.0}_{-1.0}$    	&${\phn}47.6\;^{+3.0{\phn}}_{-2.5{\phn}}$    	&${+}0.62 \pm 0.13$    	&${-}0.24 \pm 0.09$    	&${\phn}{-}10.08 \pm 1.68{\phn}$    	&main~seq.    	\\
HD~117880    	&${\phn}7914\;^{+402{\phn}}_{-360{\phn}}$    	&$2.83\;^{+0.42}_{-0.30}$    	&$1.6\;^{+1.2}_{-0.9}$    	&${\phn}14.5\;^{+2.6{\phn}}_{-2.8{\phn}}$    	&${-}2.25 \pm 0.14$    	&${-}0.27 \pm 0.45$    	&${+}145.52 \pm 3.73{\phn}$    	&HB    	\\
HD~119516    	&${\phn}5689\;^{+177{\phn}}_{-96{\phn}{\phn}}$    	&$2.23\;^{+0.33}_{-0.23}$    	&$2.8\;^{+0.2}_{-0.2}$    	&${\phn}{\phn}8.1\;^{+0.5{\phn}}_{-0.6{\phn}}$    	&${-}1.92 \pm 0.05$    	&${-}1.75 \pm 0.09$    	&${-}282.85 \pm 0.44{\phn}$    	&HB    	\\
HD~125924    	&$21898\;^{+571{\phn}}_{-938{\phn}}$    	&$4.11\;^{+0.51}_{-0.68}$    	&$0.0\;^{+2.1}_{-0.0}$    	&${\phn}68.3\;^{+7.4{\phn}}_{-6.7{\phn}}$    	&${-}0.08 \pm 0.61$    	&$< {-}0.39$    	&${+}241.09 \pm 5.86{\phn}$    	&main~seq.    	\\
HD~128801    	&$10162\;^{+291{\phn}}_{-327{\phn}}$    	&$3.54\;^{+0.36}_{-1.03}$    	&$0.0\;^{+1.3}_{-0.0}$    	&${\phn}{\phn}8.6\;^{+1.6{\phn}}_{-0.7{\phn}}$    	&${-}1.38 \pm 0.09$    	&${-}1.01 \pm 0.08$    	&${\phn}{-}79.62 \pm 2.11{\phn}$    	&HB    	\\
HD~130156    	&${\phn}6822\;^{+184{\phn}}_{-204{\phn}}$    	&$2.90\;^{+0.66}_{-0.41}$    	&$2.0\;^{+1.0}_{-1.0}$    	&${\phn}66.5\;^{+3.6{\phn}}_{-5.9{\phn}}$    	&${-}0.27 \pm 0.53$    	&$< 0.69$    	&${\phn}{-}32.49 \pm 10.02$    	&RR~Lyr ?    	\\
HD~135485    	&$15387\;^{+379{\phn}}_{-398{\phn}}$    	&$3.62\;^{+0.24}_{-0.28}$    	&$1.4\;^{+0.5}_{-0.6}$    	&${\phn}{\phn}0.0\;^{+2.1{\phn}}_{-0.0{\phn}}$    	&${+}0.40 \pm 0.14$    	&${+}0.49 \pm 0.13$    	&${\phn}{\phn}{-}3.03 \pm 0.38{\phn}$    	&poss.~HB    	\\
HD~137569    	&$12072\;^{+356{\phn}}_{-398{\phn}}$    	&$2.38\;^{+0.33}_{-0.48}$    	&$0.0\;^{+8.0}_{-0.0}$    	&${\phn}18.5\;^{+3.8{\phn}}_{-3.8{\phn}}$    	&\nodata    	&\nodata    	&${\phn}{-}59.55 \pm 1.91{\phn}$    	&poss. post-AGB    	\\
HD~143459    	&${\phn}9990\;^{+174{\phn}}_{-250{\phn}}$    	&$3.57\;^{+0.19}_{-0.29}$    	&$1.7\;^{+1.2}_{-0.9}$    	&${\phn}36.8\;^{+2.8{\phn}}_{-3.0{\phn}}$    	&${-}0.84 \pm 0.10$    	&${-}0.34 \pm 0.27$    	&${\phn}{-}19.71 \pm 2.51{\phn}$    	&HB    	\\
HD~145293    	&${\phn}6394\;^{+131{\phn}}_{-191{\phn}}$    	&$3.04\;^{+0.22}_{-0.30}$    	&$3.1\;^{+0.4}_{-0.4}$    	&${\phn}{\phn}9.8\;^{+0.9{\phn}}_{-0.8{\phn}}$    	&${-}1.11 \pm 0.11$    	&${-}0.71 \pm 0.17$    	&${\phn}{+}78.48 \pm 0.54{\phn}$    	&RR~Lyr ?    	\\
HD~159061    	&${\phn}9228\;^{+692{\phn}}_{-656{\phn}}$    	&$4.61\;^{+0.10}_{-0.23}$    	&$2.0\;^{+1.0}_{-1.0}$    	&${\phn}72.3\;^{+7.7{\phn}}_{-4.5{\phn}}$    	&${+}0.16 \pm 0.41$    	&${-}0.24 \pm 0.09$    	&${\phn}{-}14.30 \pm 4.29{\phn}$    	&subdwarf?    	\\
HD~161770    	&${\phn}5696\;^{+176{\phn}}_{-118{\phn}}$    	&$3.69\;^{+0.29}_{-0.34}$    	&$1.8\;^{+0.3}_{-0.2}$    	&${\phn}{\phn}2.6\;^{+1.2{\phn}}_{-0.9{\phn}}$    	&${-}1.81 \pm 0.09$    	&${-}1.16 \pm 0.06$    	&${-}130.13 \pm 0.55{\phn}$    	&subgiant    	\\
HD~161817    	&${\phn}7711\;^{+216{\phn}}_{-85{\phn}{\phn}}$    	&$3.22\;^{+0.18}_{-0.14}$    	&$3.9\;^{+0.4}_{-0.4}$    	&${\phn}15.2\;^{+0.8{\phn}}_{-0.7{\phn}}$    	&${-}1.52 \pm 0.05$    	&${-}1.36 \pm 0.03$    	&${-}362.65 \pm 1.29{\phn}$    	&HB    	\\
HD~166161    	&${\phn}5517\;^{+145{\phn}}_{-66{\phn}{\phn}}$    	&$2.42\;^{+0.33}_{-0.18}$    	&$2.2\;^{+0.2}_{-0.2}$    	&${\phn}{\phn}5.9\;^{+0.5{\phn}}_{-0.7{\phn}}$    	&${-}1.33 \pm 0.11$    	&${-}0.51 \pm 0.07$    	&${\phn}{+}67.85 \pm 0.21{\phn}$    	&HB    	\\
HD~167105    	&${\phn}8875\;^{+265{\phn}}_{-453{\phn}}$    	&$3.37\;^{+0.17}_{-0.16}$    	&$2.2\;^{+1.2}_{-0.8}$    	&${\phn}20.0\;^{+1.9{\phn}}_{-1.8{\phn}}$    	&${-}1.62 \pm 0.08$    	&${-}1.26 \pm 0.08$    	&${-}171.82 \pm 1.36{\phn}$    	&HB    	\\
HD~167768    	&${\phn}4823\;^{+36{\phn}{\phn}}_{-25{\phn}{\phn}}$    	&$0.82\;^{+0.18}_{-0.11}$    	&$1.7\;^{+0.2}_{-0.2}$    	&${\phn}{\phn}6.8\;^{+0.5{\phn}}_{-0.6{\phn}}$    	&${-}1.54 \pm 0.16$    	&${+}0.03 \pm 0.05$    	&${\phn}{\phn}{+}1.33 \pm 0.22{\phn}$    	&RGB    	\\
HD~170357    	&${\phn}5600\;^{+113{\phn}}_{-108{\phn}}$    	&$3.81\;^{+0.21}_{-0.30}$    	&$1.2\;^{+0.2}_{-0.2}$    	&${\phn}{\phn}3.1\;^{+1.1{\phn}}_{-0.5{\phn}}$    	&${-}0.75 \pm 0.12$    	&${-}0.17 \pm 0.04$    	&${\phn}{-}83.58 \pm 0.25{\phn}$    	&subgiant    	\\
HD~175305    	&${\phn}5149\;^{+73{\phn}{\phn}}_{-50{\phn}{\phn}}$    	&$3.23\;^{+0.27}_{-0.13}$    	&$1.5\;^{+0.2}_{-0.2}$    	&${\phn}{\phn}3.5\;^{+0.7{\phn}}_{-0.4{\phn}}$    	&${-}1.39 \pm 0.09$    	&${-}1.21 \pm 0.05$    	&${-}181.51 \pm 0.20{\phn}$    	&RGB    	\\
HD~180903    	&${\phn}7352\;^{+496{\phn}}_{-318{\phn}}$    	&$2.69\;^{+0.53}_{-0.42}$    	&$4.2\;^{+0.8}_{-0.7}$    	&${\phn}13.3\;^{+1.2{\phn}}_{-1.8{\phn}}$    	&${-}1.73 \pm 0.08$    	&${-}1.30 \pm 0.10$    	&${+}105.07 \pm 0.70{\phn}$    	&RR~Lyr ?    	\\
HD~184266    	&${\phn}5760\;^{+147{\phn}}_{-130{\phn}}$    	&$1.82\;^{+0.32}_{-0.25}$    	&$3.2\;^{+0.2}_{-0.2}$    	&${\phn}{\phn}9.3\;^{+0.8{\phn}}_{-0.6{\phn}}$    	&${-}1.73 \pm 0.11$    	&${-}0.90 \pm 0.23$    	&${-}347.54 \pm 0.32{\phn}$    	&HB    	\\
HD~195019    	&${\phn}4727\;^{+69{\phn}{\phn}}_{-62{\phn}{\phn}}$    	&$1.80\;^{+0.22}_{-0.19}$    	&$1.3\;^{+0.3}_{-0.2}$    	&${\phn}{\phn}4.3\;^{+0.7{\phn}}_{-1.5{\phn}}$    	&${-}2.32 \pm 0.09$    	&${-}0.56 \pm 0.06$    	&${-}244.36 \pm 0.30{\phn}$    	&RGB    	\\
HD~195636    	&${\phn}5399\;^{+85{\phn}{\phn}}_{-148{\phn}}$    	&$1.93\;^{+0.16}_{-0.31}$    	&$1.8\;^{+0.3}_{-0.3}$    	&${\phn}20.6\;^{+1.8{\phn}}_{-1.5{\phn}}$    	&${-}2.74 \pm 0.15$    	&${-}2.01 \pm 0.18$    	&${-}256.05 \pm 1.21{\phn}$    	&HB    	\\
HD~199854    	&${\phn}6338\;^{+158{\phn}}_{-128{\phn}}$    	&$2.18\;^{+0.29}_{-0.27}$    	&$4.6\;^{+0.5}_{-0.5}$    	&${\phn}24.3\;^{+1.6{\phn}}_{-1.6{\phn}}$    	&${-}1.71 \pm 0.08$    	&\nodata    	&${\phn}{\phn}{+}3.77 \pm 2.51{\phn}$    	&RR~Lyr ?    	\\
HD~202573    	&${\phn}4835\;^{+54{\phn}{\phn}}_{-37{\phn}{\phn}}$    	&$1.59\;^{+0.12}_{-0.16}$    	&$1.7\;^{+0.1}_{-0.1}$    	&${\phn}{\phn}6.3\;^{+0.6{\phn}}_{-0.6{\phn}}$    	&${-}1.21 \pm 0.16$    	&${-}0.14 \pm 0.05$    	&${\phn}{-}22.73 \pm 0.21{\phn}$    	&RGB    	\\
HD~203563    	&${\phn}9711\;^{+280{\phn}}_{-379{\phn}}$    	&$3.54\;^{+0.22}_{-0.44}$    	&$2.0\;^{+0.5}_{-1.0}$    	&${\phn}{\phn}3.0\;^{+1.0{\phn}}_{-1.0{\phn}}$    	&${-}0.11 \pm 0.11$    	&$< {-}4.19$    	&${\phn}{-}99.52 \pm 1.90{\phn}$    	&unknown    	\\
HD~203854    	&${\phn}5923\;^{+152{\phn}}_{-136{\phn}}$    	&$0.84\;^{+0.32}_{-0.29}$    	&$2.0\;^{+1.0}_{-1.0}$    	&$198.9\;^{+17.7}_{-14.6}$    	&${-}2.26 \pm 0.41$    	&${+}0.33 \pm 0.20$    	&${\phn}{-}48.53 \pm 24.99$    	&unknown    	\\
HD~208110    	&${\phn}5101\;^{+53{\phn}{\phn}}_{-43{\phn}{\phn}}$    	&$2.11\;^{+0.25}_{-0.21}$    	&$1.6\;^{+0.2}_{-0.2}$    	&${\phn}{\phn}4.8\;^{+0.7{\phn}}_{-0.7{\phn}}$    	&${-}1.26 \pm 0.17$    	&${-}0.42 \pm 0.06$    	&${\phn}{\phn}{-}4.25 \pm 0.20{\phn}$    	&HB    	\\
HD~210822    	&${\phn}8069\;^{+846{\phn}}_{-502{\phn}}$    	&$3.69\;^{+0.95}_{-0.80}$    	&$2.0\;^{+1.0}_{-1.0}$    	&${\phn}91.6\;^{+11.5}_{-9.4{\phn}}$    	&${-}0.26 \pm 0.27$    	&${-}0.08 \pm 0.08$    	&${\phn}{\phn}{-}2.03 \pm 7.11{\phn}$    	&main~seq.    	\\
HD~213781    	&$13322\;^{+347{\phn}}_{-374{\phn}}$    	&$3.38\;^{+0.57}_{-0.49}$    	&$0.0\;^{+2.1}_{-0.0}$    	&${\phn}34.5\;^{+4.5{\phn}}_{-3.9{\phn}}$    	&${+}0.08 \pm 0.15$    	&${-}0.76 \pm 0.28$    	&${\phn}{-}30.00 \pm 3.49{\phn}$    	&poss. HB    	\\
HD~214994    	&${\phn}9462\;^{+146{\phn}}_{-133{\phn}}$    	&$3.73\;^{+0.16}_{-0.13}$    	&$2.0\;^{+1.0}_{-1.0}$    	&${\phn}{\phn}6.7\;^{+0.5{\phn}}_{-0.6{\phn}}$    	&${+}0.09 \pm 0.05$    	&${-}0.11 \pm 0.06$    	&${\phn}{\phn}{+}9.32 \pm 0.29{\phn}$    	&main~seq. CP    	\\
HD~217515    	&${\phn}6727\;^{+394{\phn}}_{-258{\phn}}$    	&$3.79\;^{+0.50}_{-0.35}$    	&$2.0\;^{+0.3}_{-0.3}$    	&${\phn}{\phn}9.7\;^{+1.0{\phn}}_{-1.0{\phn}}$    	&${-}1.02 \pm 0.10$    	&\nodata    	&${\phn}{\phn}{-}6.25 \pm 0.41{\phn}$    	&main~seq.    	\\
HD~218790    	&${\phn}5660\;^{+157{\phn}}_{-128{\phn}}$    	&$3.41\;^{+0.36}_{-0.31}$    	&$1.0\;^{+0.2}_{-0.2}$    	&${\phn}{\phn}4.4\;^{+0.8{\phn}}_{-1.0{\phn}}$    	&${-}0.24 \pm 0.19$    	&${+}0.50 \pm 0.44$    	&${\phn}{\phn}{-}3.14 \pm 0.38{\phn}$    	&subgiant    	\\
HD~220787    	&$17747\;^{+391{\phn}}_{-699{\phn}}$    	&$3.75\;^{+0.37}_{-0.68}$    	&$2.6\;^{+5.1}_{-2.6}$    	&${\phn}26.3\;^{+3.3{\phn}}_{-3.4{\phn}}$    	&${-}0.55 \pm 0.42$    	&${-}0.33 \pm 0.08$    	&${\phn}{+}26.48 \pm 2.37{\phn}$    	&main~seq.    	\\
HD~229274    	&${\phn}5690\;^{+286{\phn}}_{-115{\phn}}$    	&$2.46\;^{+0.50}_{-0.25}$    	&$2.2\;^{+0.2}_{-0.2}$    	&${\phn}{\phn}6.7\;^{+0.6{\phn}}_{-0.6{\phn}}$    	&${-}1.40 \pm 0.10$    	&${-}0.68 \pm 0.07$    	&${-}158.64 \pm 0.22{\phn}$    	&HB    	\\
HD~233532    	&${\phn}6678\;^{+145{\phn}}_{-187{\phn}}$    	&$4.41\;^{+0.48}_{-0.42}$    	&$1.5\;^{+0.5}_{-0.5}$    	&${\phn}{\phn}9.3\;^{+0.8{\phn}}_{-0.8{\phn}}$    	&${-}0.14 \pm 0.11$    	&$< {-}1.04$    	&${\phn}{\phn}{+}3.91 \pm 0.37{\phn}$    	&main~seq.    	\\
HD~233622    	&$20595\;^{+1489}_{-2306}$    	&$3.23\;^{+1.70}_{-0.71}$    	&$2.0\;^{+1.0}_{-1.0}$    	&$276.5\;^{+24.5}_{-33.8}$    	&\nodata    	&${-}1.39 \pm 0.45$    	&${\phn}{+}31.76 \pm 22.30$    	&main~seq.    	\\
HD~233666    	&${\phn}5874\;^{+127{\phn}}_{-121{\phn}}$    	&$3.15\;^{+0.22}_{-0.21}$    	&$1.8\;^{+0.2}_{-0.2}$    	&${\phn}{\phn}5.3\;^{+0.7{\phn}}_{-0.9{\phn}}$    	&${-}1.31 \pm 0.07$    	&${-}1.47 \pm 0.10$    	&${\phn}{-}64.79 \pm 0.40{\phn}$    	&subgiant    	\\
HD~252940    	&${\phn}7652\;^{+331{\phn}}_{-204{\phn}}$    	&$3.11\;^{+0.30}_{-0.23}$    	&$3.4\;^{+0.7}_{-0.6}$    	&${\phn}22.9\;^{+2.3{\phn}}_{-1.4{\phn}}$    	&${-}1.70 \pm 0.09$    	&${-}1.50 \pm 0.07$    	&${+}161.46 \pm 1.47{\phn}$    	&HB    	\\
HZ~27    	&${\phn}9883\;^{+206{\phn}}_{-371{\phn}}$    	&$3.38\;^{+0.40}_{-0.58}$    	&$0.0\;^{+3.8}_{-0.0}$    	&${\phn}{\phn}6.6\;^{+5.0{\phn}}_{-6.6{\phn}}$    	&${-}1.39 \pm 0.35$    	&${-}0.72 \pm 0.60$    	&${\phn}{-}17.72 \pm 2.04{\phn}$    	&HB    	\\
PG0122+214    	&$25528\;^{+4448}_{-3528}$    	&$5.63\;^{+0.32}_{-1.58}$    	&$2.0\;^{+1.0}_{-1.0}$    	&$126.2\;^{+13.5}_{-6.2{\phn}}$    	&${-}0.14 \pm 0.64$    	&${-}0.20 \pm 0.15$    	&${\phn}{+}34.00 \pm 7.37{\phn}$    	&main~seq.    	\\
PG0314+146    	&${\phn}6666\;^{+513{\phn}}_{-369{\phn}}$    	&$4.48\;^{+0.62}_{-0.54}$    	&$1.2\;^{+0.8}_{-1.2}$    	&${\phn}{\phn}6.4\;^{+1.8{\phn}}_{-1.3{\phn}}$    	&${-}0.13 \pm 0.23$    	&${-}0.13 \pm 0.19$    	&${\phn}{+}13.70 \pm 0.65{\phn}$    	&main~seq.    	\\
PG0823+499    	&$16971\;^{+1214}_{-1014}$    	&$4.93\;^{+0.68}_{-0.80}$    	&$0.0\;^{+4.7}_{-0.0}$    	&${\phn}{\phn}0.0\;^{+10.9}_{-0.0{\phn}}$    	&${-}0.43 \pm 0.51$    	&${-}0.57 \pm 0.42$    	&${\phn}{+}16.23 \pm 1.78{\phn}$    	&poss. HB    	\\
PG0855+294    	&$20049\;^{+1562}_{-1502}$    	&$5.64\;^{+0.82}_{-0.91}$    	&$2.0\;^{+1.0}_{-1.0}$    	&$135.5\;^{+27.4}_{-23.0}$    	&$< {-}1.74$    	&$< 0.15$    	&${\phn}{+}61.57 \pm 31.57$    	&main~seq.    	\\
PG1205+228    	&$16271\;^{+847{\phn}}_{-1448}$    	&$3.06\;^{+1.12}_{-0.82}$    	&$2.0\;^{+1.0}_{-1.0}$    	&$175.1\;^{+28.0}_{-14.6}$    	&$< {-}2.27$    	&${-}0.63 \pm 1.23$    	&${+}139.07 \pm 20.57$    	&main~seq.    	\\
PG1530+212    	&$15000\;^{+5000}_{-5000}$    	&$4.00\;^{+2.00}_{-2.00}$    	&$2.0\;^{+1.0}_{-1.0}$    	&$103.6\;^{+16.4}_{-17.7}$    	&${-}0.05 \pm 0.74$    	&${+}0.51 \pm 0.21$    	&${\phn}{\phn}{-}7.08 \pm 24.90$    	&main~seq.    	\\
PG2219+094    	&$17402\;^{+474{\phn}}_{-3458}$    	&$3.91\;^{+0.74}_{-1.91}$    	&$2.0\;^{+1.0}_{-1.0}$    	&$225.3\;^{+28.8}_{-26.9}$    	&${-}1.88 \pm 0.65$    	&${-}1.76 \pm 0.56$    	&${\phn}{-}36.25 \pm 17.97$    	&main~seq.    	\\
PG2345+241    	&$15261\;^{+1717}_{-1453}$    	&$2.03\;^{+0.86}_{-0.35}$    	&$2.0\;^{+1.0}_{-1.0}$    	&${\phn}54.0\;^{+25.7}_{-19.9}$    	&\nodata    	&${-}0.04 \pm 0.93$    	&${\phn}{+}75.69 \pm 8.00{\phn}$    	&post-AGB    	\\
PHL~25    	&${\phn}7363\;^{+775{\phn}}_{-499{\phn}}$    	&$5.66\;^{+0.29}_{-0.38}$    	&$1.0\;^{+1.7}_{-1.0}$    	&${\phn}{\phn}6.5\;^{+3.9{\phn}}_{-4.1{\phn}}$    	&${-}0.16 \pm 0.45$    	&${-}0.32 \pm 0.37$    	&${\phn}{+}23.47 \pm 1.35{\phn}$    	&subdwarf?    	\\
PHL~3275    	&${\phn}4734\;^{+192{\phn}}_{-134{\phn}}$    	&$2.81\;^{+0.44}_{-0.45}$    	&$1.7\;^{+0.3}_{-0.3}$    	&${\phn}{\phn}2.8\;^{+1.4{\phn}}_{-2.8{\phn}}$    	&${-}0.73 \pm 0.21$    	&${+}1.51 \pm 0.74$    	&${\phn}{\phn}{-}9.56 \pm 0.27{\phn}$    	&RGB    	\\
\enddata
\end{deluxetable}

\clearpage

\begin{figure}
\epsscale{1.0}
\plotone{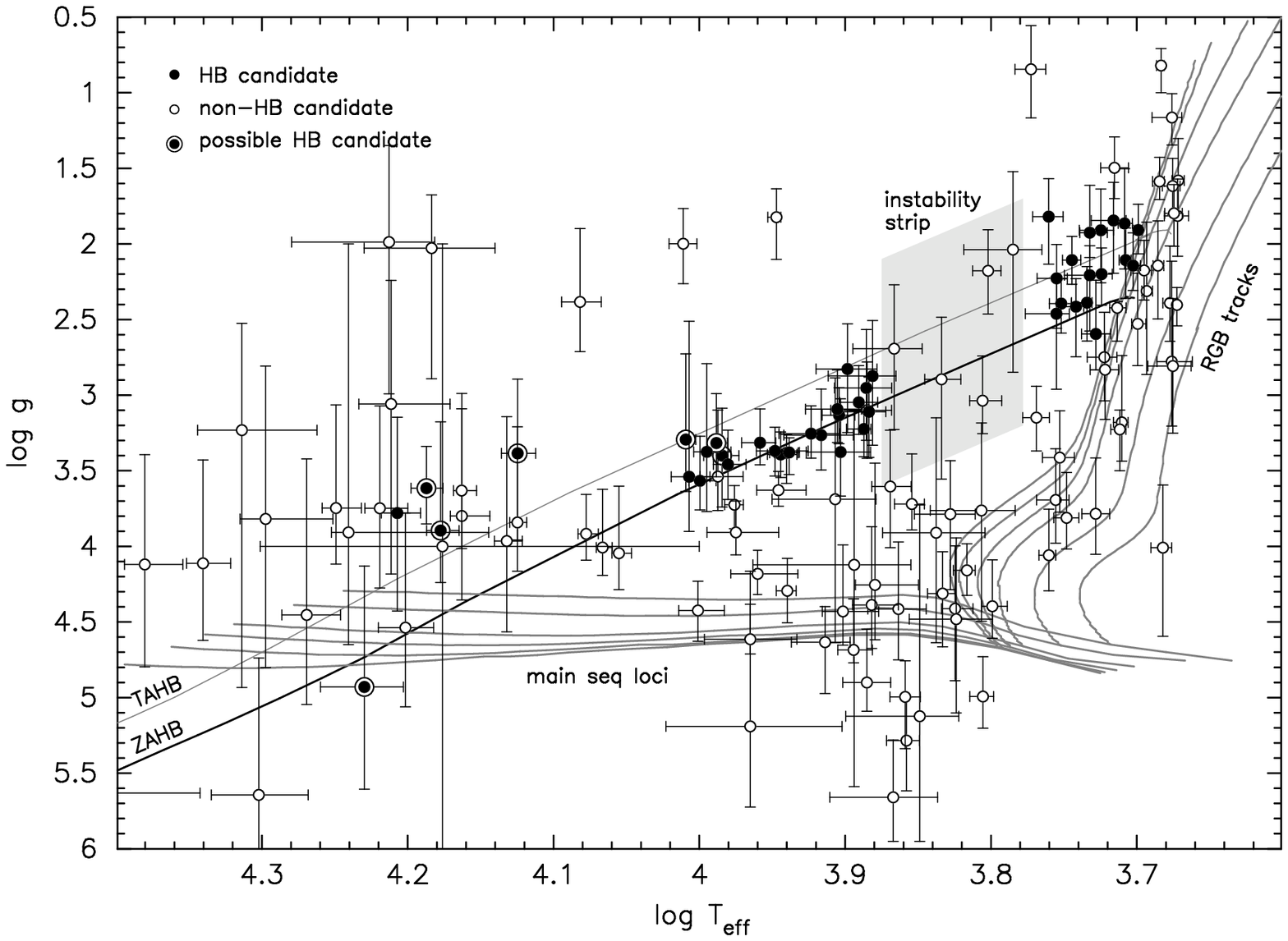}
\caption{Derived $\Teff$ and $\logg$ for program stars, plotted as an HR diagram. The zero-age horizontal branch (ZAHB) and terminal-age horizontal branch (TAHB) tracks are from \cite{dorman93}, for [Fe/H]~$=-1.48$. The RGB tracks and main sequence loci come from recent Yonsei-Yale evolution simulations \citep{yi03}. Six main sequence loci are plotted, corresponding to models with [Fe/H]$=-0.3, -0.7, -1.3, -1.7, -2.3$, and $-3.3$ (top to bottom). The RGB tracks cover the same set of metallicity values, with $M_* = 0.8 M_\odot$. \label{hr-results}}
\end{figure}

The $(\Teff, \logg)$ values for the target stars are plotted in Figure~\ref{hr-results}, in the form of an H-R diagram. Each star's position on the diagram can be compared with model tracks and loci \citep{dorman93, yi03} in order to determine its evolutionary state. Some types of stars, such as the cooler BHB stars (7500--$10000 \K$) and the subgiant branch (SGB) stars, are cleanly defined in this parameter space. In other cases, however, different stellar types occupy overlapping regions of the diagram, and it is more challenging to determine which population a specific star belongs to. Hot BHB stars and main-sequence B stars are difficult to disentangle on the basis of $\Teff$ and $\logg$ alone (particularly since the error bars are large), so in categorizing each of these stars, we consider other measured parameters, such as the abundance pattern (Population I, Population II, or metal-enhanced like hot globular cluster BHB stars) and the radial velocity (suggesting disk or halo kinematics). The RGB and the cool end of the RHB are also very close, so stellar metallicity had to be considered in these cases as well, to make sure we were comparing each star to the relevant model tracks. Our decisions regarding the nature of each target star are listed in Table~\ref{results}, and represented by different plot symbols in Figure~\ref{hr-results}.

Several of our classification choices deserve further discussion. Four of the hotter stars, with $\Teff \simeq 15000 \K$, appear to be HB objects on the basis of their small $\vsini$ and peculiar abundance patterns. PG~0823$+$499 and BD$+$30~2431 (aka Feige~86, see also \cite{bonifacio95}) both have projected rotations smaller than $10 \kms$, while BD$+$49~2137 and HD~213781 have $\vsini \simeq 33 \kms$. All four stars appear to have iron abundances near solar, somewhat lower magnesium and silicon abundances, and strong phosphorus enhancements, as listed in Table~\ref{hot-bhb-abs}. Helium is depleted in BD$+$30~2431, BD$+$49~2137, and HD~213781. Similar patterns are observed among hotter BHB stars in globular clusters \citep{glaspey89, behr99, moehler99}, and are explained as the result of chemical diffusion --- radiative levitation of most metal species, and gravitational settling of helium --- in a stable, non-convective stellar atmosphere. If the same mechanism is at work in these field stars, then the magnesium abundance is the best measure of each star's original metallicity, as the magnesium abundances in the globular cluster stars are found to correspond to the canonical cluster metallicities even when iron, phosphorus, and most other metals are strongly enhanced. We note that BD$+$49~2137 and HD~213781 rotate considerably faster than any of the metal-enhanced cluster stars, so if we can confirm that their surface abundances are due to diffusion processes, this will place useful constraints on the influence of rotation-induced mixing.

As an additional test of the nature of these four stars, we calculate their Galactic space motions. Absolute magnitudes $M_V$ are estimated from the $V$-band fluxes extracted from the model atmosphere spectral energy distributions, and a distance is computed by comparing the absolute and apparent magnitudes. Proper motion components $\mu_\alpha$ and $\mu_\delta$ are available in the Hipparcos/Tycho database for BD$+$30~2431, BD$+$49~2137, and HD~213781, while for PG~0823$+$499, we made a crude estimate of proper motion by comparing images from the first and second Palomar Sky Surveys. These data were then converted into space motion components $(U, V, W)$ using a Fortran code based on \cite{johnson87}, kindly provided by D.~Yong. (These values are computed relative to the local standard of rest, with positive $U$ values denoting motion towards the Galactic center.) According to these results, BD$+$30~2431 is clearly a member of the Galactic halo, and is therefore quite likely to be a true hot BHB star. The other stars have more disk-like velocity vectors, with the possible exception of BD$+$49~2137, so they are most likely not halo stars, but instead belong to one of the disk components of the Galaxy, which complicates the identification of their evolutionary status. As such, they are denoted as ``possible HB stars'' in Table~\ref{results}.

One other hot star, HD~135485, has also been flagged as a ``possible HB'' candidate. Its metal lines are extremely narrow, implying $\vsini < 2 \kms$, but abundance pattern is that of a metal-rich Pop~I disk star, and its space motion suggests disk membership. It shows a slight phosphorus overabundance relative to iron, but no signs of magnesium or helium depletion, so chemical diffusion is probably not active. It lies slightly above the zero-age HB locus in our HR diagram, so it may be in the process of evolving off the HB.

Three other stars with $\Teff \simeq 10000 \K$ have also been noted as ``possibles.'' Although their positions on the HR diagram appear to agree quite well with the calculated HB locus, other characteristics argue against HB membership. BD$+$30~2355 is metal-poor ([Fe{\II}/H]$\simeq -2.7$), as though it were quite old, but it rotates with $\vsini = 97.3 \kms$, and has a heliocentric $\vr$ of $1 \pm 25 \kms$, suggesting disk membership. Normally, such fast rotation would be considered ``proof'' that this not an evolved object, but since anomalously fast rotation is the primary focus of this study, we are reluctant to exclude this star from HB classification on this basis. HD~83751 and BD$-$07~0230 have disk-like radial velocities and Pop~I abundance patterns (with no sign of diffusion variations), but their rotation velocities are small: $26 \pm 1 \kms$ and $2^{+3}_{-2} \kms$, respectively. Both of these stars could be main-sequence A-type, with their polar axes nearly parallel to the line of sight (small $\sin i$), but such alignment is statistically unlikely, and would not explain their position ``above'' the main sequence. Confirmation of all three stars' low-gravity status, perhaps using spectrophotometry and Balmer line profile fitting, will probably be necessary to pin down their true nature.

\clearpage

\begin{deluxetable}{lccccc}
\tablewidth{0pt}
\tablecaption{Chemical abundances of possible hot BHB stars. \label{hot-bhb-abs}}
\tabletypesize{\small}
\tablehead{
star				&[Fe\II/H]				&[P\II/H]					&[Si\II/H]				&[Mg\II/H]				&[He\I/H]	}
\startdata
BD$+$30~2431	&$+0.38^{+0.19}_{-0.20}$	&$+1.73^{+0.09}_{-0.09}$	&$-2.85^{+0.63}_{-0.52}$	&$< -2.00 \pm 1.36$		&$-2.01^{+0.16}_{-0.18}$	\\[3pt]
HD~213781		&$+0.08^{+0.15}_{-0.15}$	&$+0.77^{+0.35}_{-0.62}$	&$-0.68^{+0.50}_{-0.65}$	&$-0.76^{+0.26}_{-0.29}$	&$-1.42^{+0.23}_{-0.27}$	\\[3pt]
BD$+$49~2137	&$+0.28^{+0.25}_{-0.49}$	&$+0.92^{+0.32}_{-0.58}$	&$-0.46^{+0.17}_{-0.19}$	&$-0.19^{+0.40}_{-0.64}$	&$-0.56^{+0.18}_{-0.19}$	\\[3pt]
PG~0823$+$499	&$-0.43^{+0.43}_{-0.58}$	&$+1.02^{+0.37}_{-0.61}$	&$+0.25^{+0.24}_{-0.30}$	&$-0.57^{+0.39}_{-0.44}$	&\nodata					\\[3pt]
HD~135485		&$+0.40^{+0.14}_{-0.14}$	&$+0.72^{+0.10}_{-0.11}$	&$+0.46^{+0.09}_{-0.10}$	&$+0.49^{+0.13}_{-0.14}$	&$+0.63^{+0.07}_{-0.08}$	\\
\enddata
\end{deluxetable}

\begin{deluxetable}{lcccccccc}
\tablewidth{0pt}
\tablecaption{Space motions of possible hot BHB stars. \label{hot-bhb-uvw}}
\tabletypesize{\scriptsize}
\tablehead{
				&			&			&heliocentric		\\[-3pt]
star				&est. $M_V$	&$d$ (pc)		&$\vr$ (km/s)	&$\mu_\alpha$ (mas/yr)	&$\mu_\delta$ (mas/yr)		&$U$ (km/s)	&$V$ (km/s)	&$W$ (km/s)		}
\startdata
BD$+$30~2431	&$+0.74$		&730		&$-26.5 \pm 0.8$		&$-17.5 \pm 1.1$\phn		&$-109.8 \pm 0.9$\phn			&$+236 \pm 23$\phn			&$-385 \pm 40$\phn		&\phn$+5 \pm 3$\phn\phn	\\
HD~213781		&$-0.07$		&660		&$-30.0 \pm 3.5$		&\phn$-0.1 \pm 1.3$\phn	&\phn\phn$+3.8 \pm 0.8$\phn	&\phn\phn$-5 \pm 4$\phn\phn	&\phn$+12 \pm 3$\phn\phn	&$+35 \pm 4$\phn\phn		\\
BD$+$49~2137	&$+0.96$		&870		&$+93.5 \pm 4.6$		&$-17.2 \pm 1.2$\phn		&\phn\phn$+2.2 \pm 1.2$\phn	&\phn$-79 \pm 8$\phn\phn		&\phn$+13 \pm 6$\phn\phn	&$+84 \pm 5$\phn\phn		\\
PG~0823$+$499	&$+3.60$		&610		&$+16.2 \pm 1.8$		&$-24.0 \pm 59.0$			&\phn$-21.0 \pm 59.0$			&\phn$-54 \pm 104$			&\phn$-34 \pm 169$		&$-41 \pm 138$			\\
HD~135485		&$+0.21$		&380		&\phn$-3.0 \pm 0.4$	&\phn$-0.4 \pm 0.9$\phn	&\phn$-33.2 \pm 0.7$\phn		&\phn$+24 \pm 2$\phn\phn		&\phn$-28 \pm 5$\phn\phn	&$-32 \pm 4$\phn\phn		\\
\enddata
\end{deluxetable}

 \clearpage
 
\begin{figure}
\epsscale{0.7}
\plotone{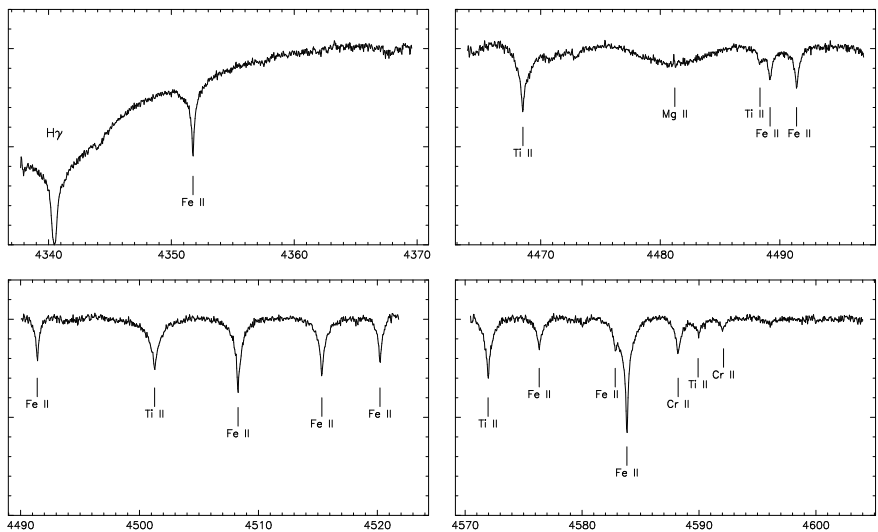}
\caption{Selected orders from the spectrum of HD~203563, an A-type field star with peculiar metal line profiles. \label{hd203563}}
\end{figure}

One other star, HD~203563, lies on the HB locus near 10000~K, but has been disqualified as a BHB star on the basis of its metal line profiles. Both \cite{stetson91} and \cite{wilhelm99} flag this object as a likely FHB candidate, and our $\Teff$ and $\logg$ values (derived from Str\"omgren photometry) would appear to support this assessment. \cite{kinman00} describe it as being ``broad-lined''. Our high-resolution spectra (Figure~\ref{hd203563}) show that the metal lines are indeed broad, but the line profiles are very much unlike those of any other stars in our sample, with narrow cores and broad wings, qualitatively similar to pressure-broadened Balmer profiles. All absorption lines are readily identified (mostly Fe{\II}, Ti{\II}, Cr{\II}, and Sc{\II}), although no line core is seen for the Mg{\II} 4481 lines, only a shallow, broad absorption feature. The H$\alpha$, H$\beta$, and H$\gamma$ line shapes appear similar to those of cool BHB stars, although the line centers reach all the way down to zero flux, which is atypical. Because the metal lines are shaped so oddly, we have been unable to run a full spectral synthesis fit, but crude manual synthesis tests suggest that [Fe/H] lies between $0.0$ and $-1.0$ dex, and $\vsini < 10 \kms$. This is certainly not a normal BHB or main-sequence star, but we have been hard-pressed to come up with an alternative identification. The line profiles imply a high-gravity photosphere, but the Tycho parallax for this star ($8.7 \pm 7.4$~mas) indicates that it is not particularly nearby, and its absolute magnitude is therefore too bright for it to be a white dwarf. Another possibility is that HD~203563 is a spectroscopic binary, with two components of similar luminosity --- one narrow-lined, one rotationally broadened by $\sim 70 \kms$ --- forming a composite line profile that merely appears to be pressure-broadened. However, the two components would have to be very similar in all other photometric parameters, in order to maintain such a similar line profile shape across all the lines, and {\it both} components would have to have zero-flux Balmer line cores, so this explanation seems unlikely. 

Any stars which lay near the HB locus, but which fell within the range $\Teff = 6000$--$7500 \K$, were considered RR~Lyrae variables, even if no photometric or $\vr$ variability was known to exist. The rotation velocities of RR~Lyraes are of interest \citep{peterson96, smith01}, but this topic lies outside the scope of the current project.


A handful of stars appear well above the HB locus in our HR diagram, and are therefore likely to be post-HB or post-asymptotic giant branch (pAGB) objects. BD+32~2188 was previously flagged as a possible pAGB star by both \cite{mitchell98} (who called it SBS10) and \cite{kinman00}, and our analysis supports this conclusion. Similarly, HD~105262 has been definitively classified as a metal-poor pAGB supergiant by \cite{reddy96}. As discussed by \citep{bolton80}, HD~137569 is tricky case, as it appears to have a faint, massive companion, perhaps a white dwarf, which could have influenced its evolution, but it is quite clearly not an HB star. At higher $\Teff$, BD+33~2642 is categorized by \cite{napiwotzki94} as the central star of a planetary nebula (CSPN) or a white dwarf, and PG~2345+241 is labeled as a ``young Population I object'' by \cite{rolleston99}, which conflicts with the low gravity that we derive on the basis of Str\"omgren photometry. We also identify HD~203854 as a pAGB object on the basis of its position in the HR diagram, although it rotates much faster than expected ($\sim 200 \kms$) for an evolved star. \cite{kun00} come to a very different conclusion for this star, identifying it as a possible young early-type dwarf with a dust disk, like Vega or $\beta$~Pictorus. Further study will be necessary to determine its evolutionary status.



\section{Reliability of $\vsini$ values}

\begin{figure}
\epsscale{0.50}
\plotone{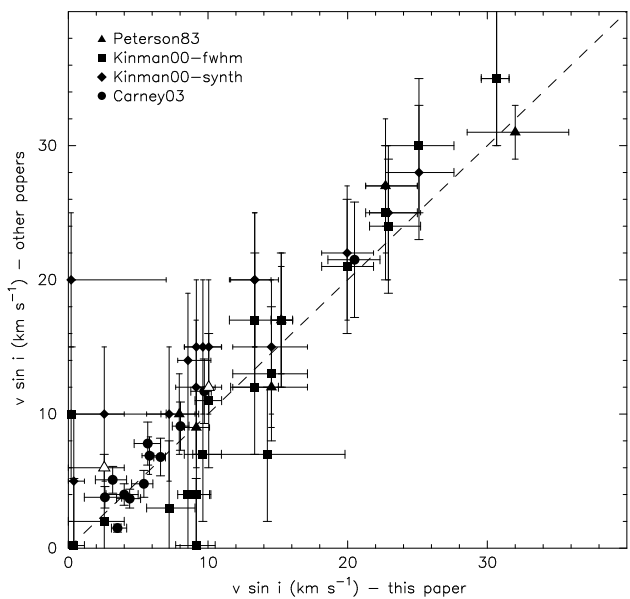}
\caption{Comparison of our $\vsini$ results to values reported in prior literature. We plot separate points for each of the two techniques used by \cite{kinman00} --- spectral synthesis fitting (``synth'', column 2 of their Table~15) and a FWHM vs. $\vsini$ calibration (``fwhm'', column 3 of their Table~15) for Mg{\II} 4481. The two open symbols denote upper bounds on the value of $\vsini$. \label{vsini-cmp}}
\end{figure}

Figure~\ref{vsini-cmp} compares our stellar $\vsini$ values to those previously reported in the literature by \cite{peterson83a}, \cite{kinman00}, and \cite{carney03}. The measurements generally agree quite well, although there are a few discrepant points, particularly in comparison to the Kinman data. In order to plot vertical error bars on all points, we estimate (somewhat arbitrary) errors of 20\% for the Carney {\etal} values, and for the Kinman results (2nd and 3rd columns of their Table~15), we adopt errors of  $\pm 5 \kms$, which is the mean difference between their two methods of measuring $\vsini$. 

The error values in Table~\ref{results} and the error bars in Figure~\ref{vsini-cmp} represent only the formal random error from the quality-of-fit analysis described in Section~3, so we also need to consider the possible influence of various other factors which could introduce systematic errors in the measurement of rotational broadening. The instrumental profile of the McDonald Cassegrain Echelle does vary as a function of position on the CCD: towards the corners, where the spectrograph camera focus degrades slightly, the instrumental profiles are approximately 10\% wider than at CCD center, so the mean instrumental profile used to broaden the synthetic spectra will be $\sim 0.25 \kms$ too wide for some spectral lines, and $\sim 0.25 \kms$ too narrow for others. Since the spectrograph is mounted on the telescope, we might also worry about flexure --- slight shifts in the instrument optics during an exposure could smear out the spectrum, making stellar absorption lines appear broader than they actually are. To address this concern, the spectrograph was built with a cantilevered counterweight system, which counteracts the gravitational pull on the main optical elements of the system. We performed several tests during cloudy nights, taking multiple thorium-argon calibration exposures over half-hour intervals as the telescope tracked, and found the wavelength shifts to be of the order of 0.1~pixels, or $0.25 \kms$, so it appears that the counterweight system is effective in minimizing flexure. There can also be a slight spectral shift during each exposure due to the Earth's rotation, but this is negligible ($< 0.1 \kms$) compared to other factors.

The rotational broadening solution will depend upon the values of $\xi$, $v_{\rm macro}$, and the limb darkening parameter $\beta_{\rm limb}$ used in the spectral synthesis analysis. With increased turbulent broadening, a smaller $\vsini$ is generally needed to match the same line profile. If $\beta_{\rm limb}$ is increased, then the ``wings'' of the rotation profile are weakened, and a larger $\vsini$ value is necessary to match the synthetic line profile with an observed profile. Furthermore, the best-fit $\vsini$ value also depends on the accuracy of the continuum normalization. The synthesis fitting routines automatically fit the continuum level on either side of each absorption line, but if the continuum is set too high or too low (due to noise, or weak lines not properly accounted for), then the line will appear too broad or too narrow, respectively.

To estimate the influence of each of these error sources on the final derived $\vsini$ values, we have selected five target stars with small to medium rotation velocities, and repeated the $\vsini$ fitting procedure under various test conditions. Table~\ref{vsini-errs} displays the results of these experiments. The first column lists the $\vsini$ result with all other parameters at their default settings, and then each subsequent column shows the change in best-fit $\vsini$ resulting from changes in the other parameters: $\xi$ increased and decreased by $1 \kms$ (a considerably larger amount than the formal error in $\xi$ for most stars), $v_{\rm macro}$ increased and decreased by $1 \kms$, limb darkening $\beta_{\rm limb}$ increased and decreased by 0.1, continuum level raised and lowered by 0.01, and the instrumental profile widened and narrowed by 5\%. In all cases, the changes in best-fit $\vsini$ due to these adjustments are smaller than the formal error bars on $\vsini$, although for some tests, the $\Delta\vsini$ is of comparable magnitude. Continuum placement errors appear to be most likely to influence the derived $\vsini$ substantially, but bear in mind that our tests artificially raise and lower the continuum for {\it all} lines simultaneously, and the actual continuum fitting errors due to spectral noise will follow a distribution of both positive and negative displacements, such that these errors will tend to cancel each other out. We estimate that all these systematic effects could increase the total errors by approximately 50\% in the worst case, so our formal error bars appear to be suitably representative of the sum of all error sources.

\clearpage

\begin{deluxetable}{lccccccccccc}
\tablecaption{Changes in best-fit $\vsini$ due to adjustment of other stellar parameters. \label{vsini-errs}}
\tabletypesize{\scriptsize}
\tablewidth{0pt}
\tablehead{
	&baseline			&$\Delta\vsini$ given:	&&&&instrumental	\\
star	&$\vsini$ (km/s)	&$\xi \pm 1$~km/s	&$v_{\rm macro} \pm 1$~km/s	&$\beta_{\rm limb} \pm 0.1$	&${\rm continuum} \pm 0.01$	&${\rm FWHM} \pm 5$\%	}
\startdata
BD+18~2890	&$\phn 3.16\;^{+0.94}_{-1.63}$	&$-0.95, -0.53$	&$-0.95, +0.32$	&$+0.01, -0.01$	&$+0.58, -0.85$	&$-0.87, +0.48$		\\[4pt]
BD+18~2757	&$\phn 5.52\;^{+0.98}_{-0.89}$	&$+0.27, -0.30$	&$-0.27, +0.25$	&$+0.02, -0.04$	&$+0.73, -0.50$	&$-0.31, +0.41$		\\[4pt]
BD+25~2602	&$13.31\;^{+1.65}_{-1.75}$	&$-0.06, -0.12$	&$-0.05, +0.02$	&$+0.02, -0.05$	&$+1.18, -0.85$	&$-0.06, +0.04$		\\[4pt]
HD~109995	&$22.87\;^{+2.10}_{-	1.57}$	&$+0.92, -0.63$	&$-0.02, +0.11$	&$+0.16, -0.09$	&$+1.69, -1.17$	&$-0.05, +0.07$		\\[4pt]
BD+46~1998   	&$31.80\;^{+1.51}_{-1.52}$	&$+1.14, -0.51$	&$+0.04, -0.01$	&$+0.12, -0.10$	&$+1.32, -1.44$	&$-0.11, -0.01$		\\[4pt]
\enddata
\end{deluxetable}

\clearpage


\section{Trends in rotation velocity}
	
\begin{figure}
\epsscale{0.70}
\plotone{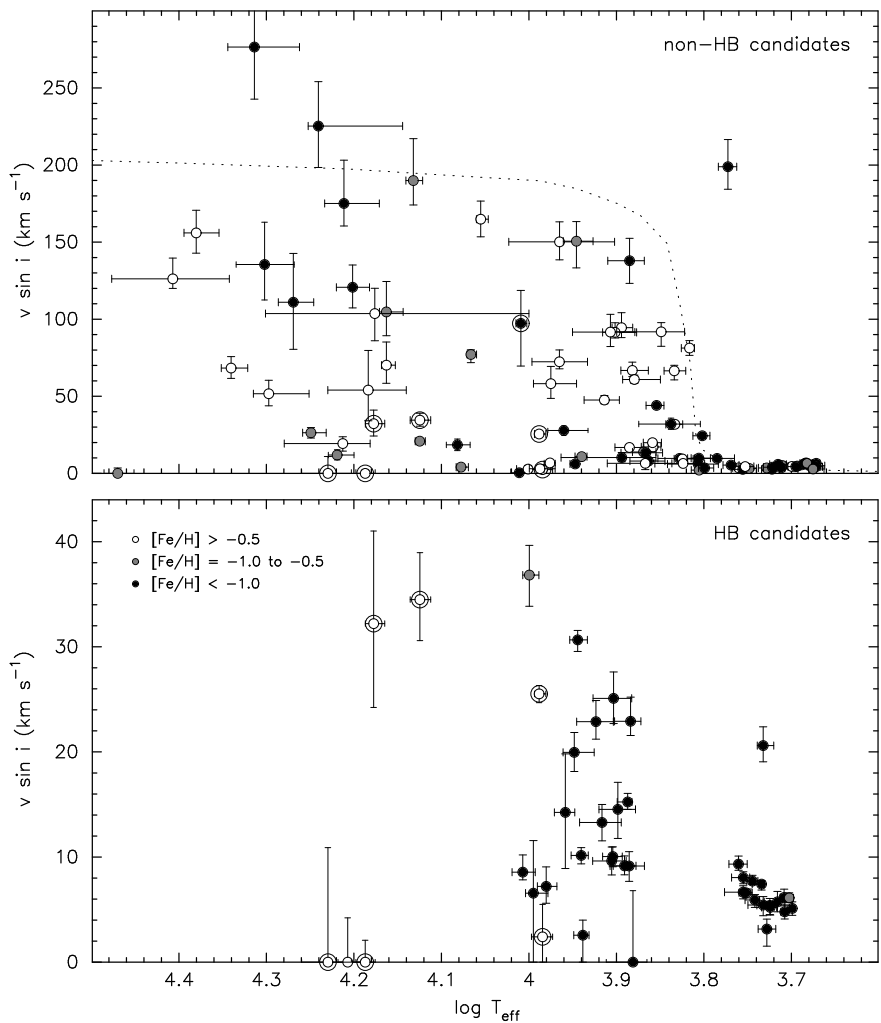}
\caption{Projected rotation velocity $\vsini$ as a function of $\Teff$. The bottom panel shows stars which were chosen as likely RHB or BHB stars, while the top panel (with a different vertical scale) shows stars which are probably {\it not} HB stars. The dashed line in the upper panel denotes the expected magnitude of stellar rotation for main sequence stars. The four stars whose classification is uncertain are plotted in both panels (although one of them does not fit within the vertical scale of the bottom panel), and are flagged with circles around the plot symbols. \label{vsini-teff}}
\end{figure}

Figure~\ref{vsini-teff} plots the measured values of $\vsini$ as a function of the derived effective temperature for each star. In the upper panel, the non-HB stars appear to follow the expected distribution of $\vsini$ quite closely --- nearly all stars lie below the dashed line, which shows the average $\vrot$ for main sequence stars of various spectral types, as quoted by \cite{allen00}. One notable exception is HD~203854, the cool low-gravity star with $\vsini \simeq 200 \kms$ that was discussed previously.

In the lower panel of Figure~\ref{vsini-teff}, we plot the $\vsini$ values for stars identified as HB candidates. We will separately consider three different temperature regimes along the HB locus: the RHB stars with $\Teff = 4900$--$5900 \K$ ($\log\Teff = 3.69$--3.77), the cool BHB stars with $\Teff = 7500$--$11500 \K$ ($\log\Teff = 3.88$--4.06), and the hot BHB stars with $\Teff > 11500 \K$ ($\log\Teff > 4.06$). The dividing line between cool and hot BHB stars is chosen to coincide with the location of the photometric ``jumps'' \citep{grundahl99} found in globular cluster BHB color-magnitude diagrams, which appear to be related to an abrupt change in stellar rotation and abundance characteristics at this threshold temperature.

Among the hot BHB population with $\Teff > 11500 \K$, three of the stars have zero or very small rotational broadening, much like the analogous hot HB stars in globular clusters \citep{behr00a, recioblanco02}, although the two other hot field stars exhibit substantial rotation, $\vsini > 30 \kms$. All five stars show metal enhancement patterns similar to those seen among hotter BHB stars in globular clusters. These abundance variations are attributed to radiative levitation and chemical diffusion mechanisms \citep{michaud83, glaspey89}, which can operate only in the absence of convection or other sources of mixing, such as the meridional circulation induced by fast stellar rotation. If a star is rotating faster than some threshold velocity, then the induced circulation currents should prevent the metal enhancements from appearing. On the basis of two hot but metal-poor (unenhanced) BHB stars in M15, \cite{behr00b} claimed that this threshold lay at $\vrot \simeq 10 \kms$, but recent theoretical calculations by \cite{michaud03} suggest that the threshold velocity could be higher. The existence of these two hot stars with metal enhancements {\it and} fast rotation would appear to provide further evidence in favor of a higher $\vrot$ threshold, if they are indeed BHB stars.

The cool BHB field stars span a wide range of $\vsini$, as was first discovered by \cite{peterson83a}, and corroborated by \cite{kinman00}. The observed distribution of $\vsini$ values suggests that the underlying distribution of true rotation velocities $\vrot$ is roughly bimodal, with a fast population of $\vrot \simeq 30$--$35 \kms$, and a slower population of $\vrot \simeq 10$--$15 \kms$, similar to the bimodal populations found in many globular clusters. We had hoped, with this project, to substantially expand the list of known field BHB stars in this temperature range, in order to better constrain the underlying distribution $\vrot$, and assess its similarities to the bimodal distributions found in globular clusters, but with only five new cool BHB stars added to the sample, we can offer no substantive new insights regarding the $\vrot$ distribution for the field population.


With one exception, all of the RHB stars in our sample have $\vsini < 10 \kms$, and there appears to be some correlation between rotation velocity and effective temperature, with hotter stars tending to have higher $\vsini$. A trend of this sort would be expected even if all these stars had the same total angular momentum, given that stellar mass and radius both decrease rapidly with increasing surface temperature along this part of the HB. (Very crudely, since moment of inertia $I \sim R^2$ for a sphere of uniform density, a constant $L$ implies $\omega \sim R^{-2}$, and $\vrot \sim \omega R \sim R^{-1}$, so hotter stars, having smaller $R$, should show higher $\vrot$.) Detailed modeling of the internal density structure of these stars (along the lines of Table 1 of \cite{sills00}) will be necessary to accurately estimate how moment of inertia varies as a function of position along the HB locus, so that we can see whether these RHB stars all have similar amounts of angular momentum, and determine how the RHB stars compare to the BHB population.

The one outlier among the RHB stars, HD~195636, has a significantly higher projected rotation velocity of $\sim 20 \kms$, as noted previously by \cite{preston97} and \cite{carney03}. This star (along with three other RHB stars with $\vsini > 10 \kms$ observed by Carney \etal, but not in our sample) might represent a fast-rotating population analogous to that seen among the cool BHB stars. Observations of additional RHB stars, both in the field and in globular clusters, will be useful for hunting down the extent and origin of the fast-rotating population.

\begin{figure}
\epsscale{0.50}
\plotone{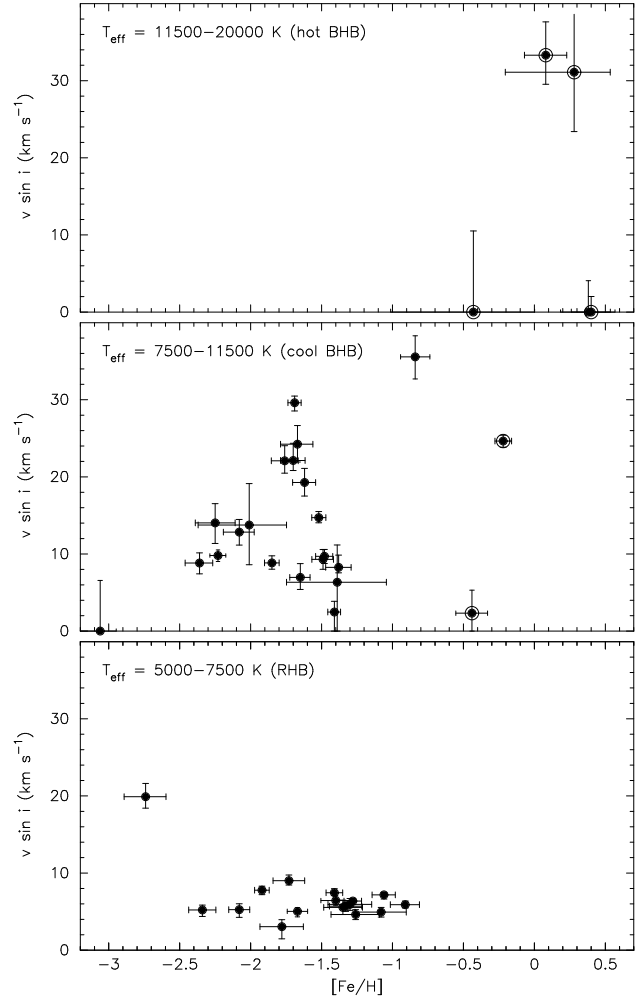}
\caption{Projected rotation velocity $\vsini$ as a function of derived stellar metallicity. The top panels shows the hottest BHB stars, the middle panel shows the cooler BHB stars, and the bottom panel shows the RHB stars. \label{vsini-metal}}
\end{figure}

Figure~\ref{vsini-metal} displays the same HB $\vsini$ values as a function of derived stellar metallicity. Stars are grouped into the same three different temperature ranges described above. No clear statistical correlation between $\vsini$ and [Fe/H] is evident, although we note that many of the fast-rotating cool BHB stars cluster around ${\rm [Fe/H]} \simeq -1.7$ (see also Figure~12a of \cite{kinman00}), and the fastest-rotating RHB star other than HD~195636 also lies near this metallicity. This ``peak'' in $\vsini$ may be merely a result of small-number statistics, but if a correlation of this sort were actually present, it would provide a vital clue regarding the origin of the fast rotation, so future surveys of BHB rotation velocities should be sure to thoroughly sample the entire range ${\rm [Fe/H]} = -2.5$ to $-1.0$, in order to test the possibility that there is something ``special'' about ${\rm [Fe/H]} \simeq -1.7$.



The presence of fast-rotating cool BHB stars in the field population would seem to argue against the idea that fast rotation is a signature of dense stellar environments, as has been suggested for the fast rotators in globular clusters. Within a cluster, an appreciable fraction of stars might be expected to have had close tidal ``fly-by'' encounters with other cluster stars, which could impart additional angular momentum to one or both stars. In the halo, however, the probability of such interactions is exceedingly low. It would be difficult to argue that close tidal encounters are responsible for the fast rotation among field HB stars, unless one is willing to accept that many of the ``halo'' HB stars originated in globular clusters, and were then stripped away by the tidal gravitational influence of the Galaxy. Since the fraction of field BHB stars with fast rotation ($\sim 0.3$) is comparable to the fraction of fast rotators in some globular clusters ($\sim 0.4$ for M13 and M68, $\sim 0.3$ for M15 and M92), we must tentatively conclude that the surrounding dynamical environment does not influence stellar rotation to any significant degree.

To fully test this hypothesis, we would like to subdivide the FHB stars into many subpopulations, on the basis of metallicity, Galactic position, and full 3-D space motions, so we could then deduce the $\vrot$ distribution of each subpopulation separately, and see whether fast rotation can be linked to some other specific stellar property. However, our current sample of confirmed FHB stars is not nearly large enough for such partitioning. An analysis of this sort will have to await a more extensive future $\vsini$ survey, using a large telescope to measure the abundances and line broadening of many hundreds of fainter field HB candidates identified by spectrophotometry \citep{wilhelm99} or multicolor photometry \citep{yanny00}.

	
\section{Main sequence A-star contaminants}

A large fraction of the stars observed during this project turn out to be rather ordinary fast-rotating B and A dwarfs. The selection criteria used to identify target stars were defined quite broadly (so as not to accidentally exclude any interesting or peculiar stars), so it would not be surprising to find that a few ``contaminants'' (non-evolved Population~I stars) had snuck into our target list. However, since many of the catalogs that we used for target selection were designed to find halo stars, it seems mildly puzzling that so many apparently ordinary A and B stars have turned up in our sample. For instance, we expected to find a high fraction of HB candidates among the faint blue high-latitude stars listed by \cite{newell73} and \cite{newell76}, especially in light of the ``gaps'' in the color distribution of these stars (see Figure 1 of both papers), which appeared similar to the gaps found along the HBs of globular clusters. But fewer than half (11 of 27) of the Newell stars that we observed were clearly HB objects, with another 11 stars classified as Population~I dwarfs, and the remaining 5 stars marked as pAGB, subgiants, and such. This raises two questions: how did such a large fraction of non-evolved stars enter this sample, and why would gaps appear in the color distribution of such a heterogenous set of stars?

A partial answer to the first question may come from recent studies of early-type dwarfs which are found far from the Galactic plane, as though members of the halo population. Detailed kinematic and abundance analyses, such as those described by \cite{rolleston99} and \cite{ramspeck01}, suggest that some faint blue high-latitude stars are formed within the disk but then thrown out into the halo by supernova kicks or dynamical encounters within their birth clusters. In our search for HB stars, we seem to have stumbled across several of these ``runaway B stars,'' described in earlier papers:  HD~125924 \citep{quin91}, HD~100340 \citep{ryans99}, PG~1205+228 \citep{conlon89, conlon90, saffer97}, and Feige~84 \citep{greenstein74}. Feige~40 may also belong to this category.

We have also ``discovered'' three chemically peculiar (CP) A-type stars, which show enhanced abundances of manganese and other selected metals. All three of these stars have been previously identified as being chemically peculiar --- HD~214994 (omicron Pegasi) by \cite{adelman88}, HD~27295 (53 Tau) by \cite{adelman87}, and HD~7374 by \cite{jomaron99} --- and as expected, all three stars have small rotation velocities compared to the typical main sequence A star.


\section{Conclusion}

We have made high-resolution spectroscopic observations of a large sample of brighter ($V \simeq 6$--11) field stars, in order to identify horizontal branch stars and determine their projected rotation velocities via spectral synthesis fitting. We present $\vsini$ measurements for 45 HB stars, of which 22 have no prior rotation measurement reported in the literature. The bimodal distribution of rotation velocities that we find among field BHB stars is similar to that of globular cluster BHB stars, suggesting that the cluster environment is not responsible for the anomalously fast rotation exhibited by some HB stars. Our field RHB stars do not show any obvious signs of an analogous bimodality in $\vrot$, although the possibility cannot be ruled out, in light of the RHB measurements of \cite{carney03}. A few hotter BHB stars exhibit abundance patterns suggestive of radiative levitation of metals and gravitational settling of helium, which may occur even when a star is rotating at $\vrot \simeq 30 \kms$.

Future progress towards an understanding of the rotation characteristics of HB stars (and the larger issue of how a low-mass star's angular momentum changes as it evolves) will require larger samples, with good coverage throughout parameter space ($\Teff$, [Fe/H], and $(U, V, W)$ space motions), so that we can disentangle the potential differences among kinematic populations (disk, halo, and even velocity substructures within the halo), uncover the dependence (if any) with metallicity, and fully map out the underlying distribution of rotation rates of the different populations along the HB locus. Observing a sufficiently large sample of field HB stars at the requisite spectral resolution and $S/N$ ratio will probably entail an extensive program on a 6 to 8-meter class telescope, in order to reach significantly fainter FHB stars ($V \sim 10-14$) than have been observed to date.


\clearpage
\acknowledgements

I am grateful to the staff of McDonald Observatory for their support of the observations, and also I thank the McDonald TAC for the extensive observing time allocated for this project. The HIRES spectra of selected faint targets were collected during Keck time graciously contributed by Judy Cohen and Jim McCarthy; the W.M. Keck Observatory is operated jointly by the California Institute of Technology and the University of California. I received several helpful suggestions from Inese Ivans and Chris Sneden (regarding the spectral synthesis fitting procedures) and from the anonymous referee (regarding the kinematic characteristics of the hot BHB star candidates). Thanks also go to Robert Kurucz, Michael Lemke, David Yong, and the VALD team for making their computer codes and data sets available.


\clearpage

\end{document}